\documentclass[12pt]{article}
\usepackage{cite}
 \advance\oddsidemargin by -1in
 \advance\evensidemargin
 by -1in
 \marginparsep
 0.4cm
 \advance\topmargin by
 -0.0in
 \makeindex
 \def\gsim{\mathrel{\rlap{\lower4pt\hbox{\hskip1pt$\sim$}}
 \raise1pt\hbox{$>$}}}

 \newcommand\la{\langle}
 \newcommand\ra{\rangle}
 \newcommand\beq{\begin{equation}}
 \newcommand\noi{\noindent}
 \newcommand\eeq{\end{equation}}
 \newcommand\beqn{\begin{eqnarray}}
 \newcommand\eeqn{\end{eqnarray}}
\def\lsim{\mathrel{\rlap{\lower4pt\hbox{\hskip1pt$\sim$}}
    \raise1pt\hbox{$<$}}}         
\def\gsim{\mathrel{\rlap{\lower4pt\hbox{\hskip1pt$\sim$}}
    \raise1pt\hbox{$>$}}}         
 \newcommand{\doublespace} {
\def\Re{\,\mbox{Re}\,}
\def\Im{\,\mbox{Im}\,}
\def\mb{\,\mbox{mb}}
\def\fm{\,\mbox{fm}}  
\def\GeV{\,\mbox{GeV}}
\def\MeV{\,\mbox{MeV}}
\def\sig{\sigma_{\bar qq}^N}
 \renewcommand{\baselinestretch} {1.6}
 \large\normalsize}

\begin{document}

\title{\hspace*{10cm} {\large NSF-ITP-02-40\\
\hspace*{9cm} ITP, UCSB,
Santa Barbara}\\[50pt]
\bf Gluon Shadowing in Heavy Flavor 
Production off Nuclei}
 
\maketitle
 
\begin{center}
 
 {\large B.Z.~Kopeliovich$^{1-3}$ and A.V.~Tarasov$^{3}$ 
}
 \\[1cm]
 $^{1}${\sl Max-Planck Institut f\"ur Kernphysik, Postfach 103980, 69029
Heidelberg}\\[0.2cm]
 $^{2}${\sl Institut f\"ur Theoretische Physik der Universit\"at, 93040
Regensburg} \\[0.2cm]
 $^{3}${\sl Joint Institute for Nuclear Research, Dubna, 141980 Moscow
Region, Russia}\\[0.2cm]

\end{center}                                                              
 
\vspace{1cm}
 
\begin{abstract} \noi Gluon shadowing which is the main source of nuclear effects for
production of heavy flavored hadrons, remains unknown. We develop a light-cone dipole
approach aiming at simplifying the calculations of nuclear shadowing for heavy flavor
production, as well as the cross section which does not need next-to-leading and
higher order corrections. A substantial process dependence of gluon shadowing is found
at the scale of charm mass manifesting a deviation from QCD factorization. The
magnitude of the shadowing effect correlates with the symmetry properties and color
state of the produced $\bar cc$ pair. It is about twice as large as in DIS
\cite{kst2}, but smaller than for charmonium production \cite{kth}. The higher twist
shadowing correction related to a nonzero size of the $\bar cc$ pair is not negligible
and steeply rises with energy.  We predict an appreciable suppression by shadowing for
charm production in heavy ion collisions at RHIC and a stronger effect at LHC.  At the
same time, we expect no visible difference between nuclear effects for minimal bias
and central collisions, as is suggested by recent data from the PHENIX experiment
\cite{phenix-charm}. We also demonstrate that at medium high energies when no
shadowing is possible, final state interaction may cause a rather strong absorption of
heavy flavored hadrons produced at large $x_F$.

\end{abstract}

\doublespace

\newpage
 
\vspace*{1cm}

\section{Introduction}

It is still unclear whether available data from fixed target experiments
demonstrate any nuclear effects for open charm production
\cite{mike,e769,wa82}.  Naively one might expect no effects at all, since a
heavy quark should escape the nucleus without attenuation or reduction of its
momentum. In fact, this is not correct even at low energies as is explained
below.  Moreover, at high energies one cannot specify any more initial or
final state interactions. The process of heavy flavor production takes a time
interval longer than the nuclear size, and the heavy quarks are produced
coherently by many nucleons which compete with each other.  As a result the
cross section is reduced, and this phenomenon is called shadowing. 

In terms of parton model the same effect is interpreted in the infinite
momentum frame of the nucleus as reduction of the nuclear parton density
due to overlap and fusion of partons at small Bjorken $x$. The
kinematic condition for overlap is the same as for coherence in the
nuclear rest frame. 

There are well known examples of shadowing observed in hard reactions, like
deep-inelastic scattering (DIS) \cite{nmc} and Drell-Yan process (DY) 
\cite{e772} demonstrating a sizeable reduction of the density of light sea
quarks in nuclei. Shadowing is expected also for gluons, although there is
still no experimental evidence for that.

Shadowing for heavy quarks is a higher twist effect, and although its
magnitude is unknown within the standard parton model approach, usually it is
neglected for charm and beauty production. However, this correction is
proportional to the gluon density in the proton and steeply rises with
energy. Unavoidably, such a correction should become large at high energies. 
In some instances, like for charmonium production, this higher twist effect
gains a large numerical factor and leads to a rather strong suppression even
at energies of fixed target experiments \cite{kth}.

On the other hand, gluon shadowing which is the leading twist effect, is
expected to be the main source of nuclear suppression for heavy flavor
production at high energies.  This is why this process is usually
considered as a sensitive probe for the gluon density in hadrons and
nuclei.  If to neglect the power, $1/m_Q^2$, corrections (unless otherwise
specified), then the cross section of heavy $\bar QQ$ production in $pA$
collision is suppressed by the gluon shadowing factor $R^G_A$ compared to
the sum of $A$ nucleon cross sections,
 \beq
\sigma^{\bar QQ}_{pA}(x_1,x_2) = 
R^G_A(x_1,x_2)\, A\,\sigma^{\bar QQ}_{pN}(x_1,x_2)\ .
\label{2}
 \eeq
 Here 
 \beq 
R^G_A(x_1,x_2) = {1\over A}\int d^2b\,R^G_A(x_1,x_2,b)\,T_A(b)\ , 
\label{3}
 \eeq   
 where $R^G_A(x_1,x_2,b)$ is the (dimensional) gluon shadowing factor at
impact parameter $b$;  $T_A(b) = \int_{-\infty}^{\infty}\,dz\,\rho_A(b,z)$ is
the nuclear thickness function at impact parameter $b$;  $x_1,\ x_2$ are the
Bjorken variables of the gluons participating in $\bar QQ$ production from
the colliding proton and nucleus. 

Parton model cannot predict shadowing, but only its evolution at high $Q^2$,
while the main contribution originates from the soft part of the interaction. 
The usual approach is to fit data at different values of $x$ and $Q^2$
employing the DGLAP evolution and fitting the distributions of different
parton species parametrized at some intermediate scale \cite{eks,kumano}.
However, the present accuracy of data for DIS on nuclei do not allow to fix
the magnitude of gluon shadowing, which is found to be compatible with zero
\footnote{Gluon shadowing was guessed in \cite{eks} to be the same as for
$F_2(x,Q^2)$ at the semi-hard scale.}. Nevertheless, the data exclude some
models with too strong gluon shadowing \cite{eks-new}. 

Another problem faced by the parton model is impossibility to predict gluon
shadowing effect in nucleus-nucleus collisions even if the shadowing factor
Eq.~(\ref{2}) in each of the two nuclei was known. Indeed, the cross section
of $\bar QQ$ production in collision of nuclei $A$ and $B$ at impact
parameter $\vec b$ reads,
 \beq
\frac{d\sigma^{\bar QQ}_{AB}(x_1,x_2)}{d^2b} = 
R^G_{AB}(x_1,x_2,b)\,AB\,
\sigma^{\bar QQ}_{NN}(x_1,x_2)\ ,
\label{4}
 \eeq
 where 
 \beq
R_{AB}^G(x_1,x_2,b) = \frac{1}{AB}
\int d^2s\,R^G_A(x_1,\vec s)\,T_A(s)\ 
R^G_B(x_2,\vec b-\vec s)\,T_B(\vec b-\vec s)\ .
\label{6}
 \eeq
 In order to calculate the nuclear suppression factor Eq.~(\ref{6})  one
needs to know the impact parameter dependence of gluon shadowing,
$R^G_A(x_1,\vec b)$, while only integrated nuclear shadowing Eq.~(\ref{3}) 
can be extracted from lepton- or hadron-nucleus data\footnote{One can get
information on the impact parameter of particle-nucleus collision measuring
multiplicity of produced particles or low energy protons (so called grey
tracks). However, this is still a challenge for experiment.}. Note that
parton model prediction of shadowing effects for minimum bias events
integrated over $b$ suffers the same problem.  Apparently, QCD factorization
cannot be applied to heavy ion collisions even at large scale. The same is
true for quark shadowing expected for Drell-Yan process in heavy ion
collisions \cite{hir,krtj,gay}.
            
Nuclear shadowing can be predicted within the light-cone (LC) dipole approach
which describes it via simple eikonalization of the dipole cross section. It
was pointed out in \cite{zkl} that quark configurations (dipoles) with fixed
transverse separations are the eigenstates of interaction in QCD, therefore
eikonalization is an exact procedure. In this way one effectively sums up the
Gribov's inelastic corrections in all orders \cite{zkl}. 

The advantage of this formalism is that it does not need any $K$-factor,
since effectively includes all higher twist and next-to-leading,
next-to-next... corrections. Of course, one assumes universality of the
dipole cross section, i.e. its process independence. This property can be
proven in the leading $\log(1/x)$ approximation, otherwise is an assumption
of the model. Indeed, it was demonstrated recently in \cite{joerg} that the
simple dipole formalism for Drell-Yan process \cite{hir,bhq,kst1} precisely
reproduces the results of very complicated next-to-leading calculations at
small $x$.  The LC dipole approach also allows to keep under control
deviations from QCD factorization.  In particular, we found a substantial
process-dependence of gluon shadowing due to existence of a semi-hard scale
imposed by the strong nonperturbative interaction of light-cone gluons
\cite{kst2}. For instance gluon shadowing for charmonium production off
nuclei was found in \cite{kth} to be much stronger than in deep-inelastic
scattering \cite{kst2}. 

The LC dipole approach also provides effective tools for calculation of
transverse momentum distribution of heavy quarks, like it was done for
radiated gluons in \cite{kst1,jkt}, or Drell-Yan pairs in \cite{krtj}.
Nuclear broadening of transverse momenta of the heavy quarks also is an
effective way to access the nuclear modification of the transverse momentum
distribution of gluons, so called phenomenon of color glass condensate or
gluon saturation \cite{mv,al}. In this paper we consider only integrated
quantities and leave the transverse momentum distribution for further
calculations. 

In what follows we find sizeable deviations from QCD factorization for heavy
quark production off nuclei. First of all, for open charm production
shadowing related to propagation of a $\bar cc$ pair through a nucleus is not
negligible, especially at the high energies of RHIC and LHC, in spite of
smallness of $\bar cc$ dipoles. Further, higher Fock components containing
gluons lead to gluon shadowing which also deviates from factorization and
depends on quantum numbers of the produced heavy pair $\bar cc$.

This paper is organized as follows. In Sect.~\ref{NN} we consider the
contribution of the lowest Fock component $\bar cc$ of projectile gluons to
production of a $\bar cc$ pair in $NN$ collisions.  A LC representation is
derived for three production amplitudes which are classified in accordance
with the color and symmetry properties of the $\bar cc$. We are focusing on
charm for the sake of concreteness, but all the results of the paper are
applicable to production of beauty. 

Multiple color-exchange interaction of a $\bar cc$ pair propagating through a
nucleus are considered in Sect.~\ref{eikonal}. The corresponding nuclear
shadowing has a form of the conventional Glauber eikonal, where the exponent
contains the dipole cross section $\sigma_3$ of interaction of a 3-parton
system $\bar ccG$. 

Sect.~\ref{shadowing} is devoted to the problem of gluon shadowing.  First of
all, a dipole representation for the process $GN\to \bar cc\,G\,X$ is derived
in Sect.~\ref{gl-rad} and \ref{diagrams}. Three types of amplitudes are
found, one for production of a colorless $\bar cc$ and a gluon, and two for
color-octet $\bar cc$ and a gluon, corresponding to two possible symmetries
of an octet-octet state.

The same reaction of gluon radiation accompanying $\bar cc$ pair production
off nuclei is related to the phenomenon of gluon shadowing. It is described
in Sect.~\ref{gl-rad-nucl} employing the technique of light-cone Green
functions corresponding to propagation of a $\bar ccG$ system through nuclear
matter.

Light-cone gluons are known to have a short range propagation radius in the
transverse plane which we describe via a nonperturbative light-cone potential
in the Schr\"odinger two-dimensional equation for the Green function.  The
strength of the potential depends on the color state and symmetry of the
$\bar ccG$ system.  Correspondingly, the LC nonperturbative wave functions
are also different.  In Sect.~\ref{nonpqcd} we find three types of LC wave
functions with different mean separations. 

Naturally, gluon shadowing correlates with the size of the $\bar ccG$,
therefore it demonstrates a strong process dependence, i.e.  deviation from
QCD factorization. Different types of gluon shadowing calculated in
Sect.~\ref{process-dep} are presented in Fig.~\ref{gl-shad}.  Since about
$20\%$ of $\bar cc$ pair are produced in a colorless state, it is important
to know the fraction of such pairs which end up in an open charm channel. 
This estimate is done in Sect,~\ref{singlet}.

The results of these calculations are used in Sect.~\ref{charm-shad} to
predict nuclear shadowing for open charm production in proton-nucleus
collisions. This includes gluon as well as charm quark shadowing, and also
possible medium modifications like EMC effect and gluon antishadowing.  We
provide predictions of nuclear effects plotted in Fig.~\ref{pa-xf} for open
charm production in $p-W$ collisions at the energy of the HERA-B experiment. 

In the same Sect.~\ref{charm-shad} we provide predictions for shadowing for
charm production in heavy ion collisions at the energies of RHIC
($\sqrt{s}=200\GeV$) and LHC ($\sqrt{s}=5500\GeV$) depicted in
Figs.~\ref{aa-mb} -- \ref{aa-centr}. We found quite a sizeable contribution
from the higher twist effect of shadowing related to size of the $\bar cc$
pair. A most interesting observation is nearly identical shadowing effects
predicted for minimal bias and central collisions, what has been indeed
observed recently by the PHENIX experiment at RHIC. We identify the source of
such a coincidence, and emphasize that this observation should not be
interpreted as an indication for weak nuclear effects.  Indeed,
Figs.~\ref{aa-centr} demonstrate a substantial nuclear shadowing even for
RHIC.

In Sect.~\ref{low-energy} we consider the case of medium high energies when
noshadowing is possible since the coherence length is short.  Then the $\bar
cc$ pair is produced momentarily on a bound nucleon and then undergoes final
state interactions.  On the contrary to wide spread believe, we argue that
these interactions lead to absorption related to an unusual configuration in
which the heavy flavored hadron is created. 

We summarize the results of calculations and observations in the concluding
Sect.~\ref{summary}. 

\section{Light-cone dipole formalism for charm production}\label{LC}

\subsection{\boldmath$NN$ collisions}\label{NN}
 
\begin{figure}[tbh]
\includegraphics{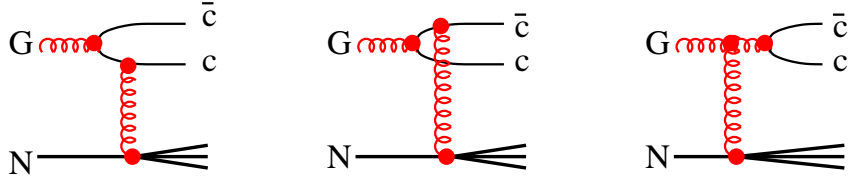} 
\begin{center} 
\vspace{4cm} 
\parbox{13cm}
 {\caption[Delta]
 {\sl Perturbative QCD mechanism of charm production production in a
gluon-nucleon collision. Only the lowest $\bar cc$ Fock component of
the gluon is taken into account.}
 \label{3graphs}}
\end{center}
 \end{figure}

For the sake of concreteness in what follows we consider charm $\bar cc$ pair
production, unless otherwise specified. Our results are easily generalized to
the case of heavier quarks. The parton model treats this process in the rest
frame of the produced pair as glue-glue fusion, $GG\to\bar cc$. At the same
time, in the rest frame of the nucleus it looks like interaction of a $\bar
cc$ fluctuation which has emerged from a projectile gluon. Thus, the problem
is reduced to the process,
 \beq
G + N \to \bar cc + X\ .
\label{100}
 \eeq 
 In the LC dipole approach the cross section is represented by a sum over
different Fock components of the projectile gluon whose LC wave functions
squared are
convoluted with proper dipole cross sections.  The cross section
corresponding to Feynman graphs depicted in Fig.~\ref{3graphs} was calculated
in \cite{npz} and it was found that it needs a dipole cross section
corresponding to a three-body system $G\bar cc$.
 This observation follows the general prescription \cite{hir} that the
partonic process $a\to bc$ is related to the dipole cross section
$\sigma_{\bar abc}(\vec r,\alpha)$ for the three-partonic $\bar abc$ system
with transverse separations $\vec r$ (for the $bc$), $\alpha\vec r$ ($\bar
ac$) and $\bar\alpha\vec r$ ($b\bar a$). Here $\alpha$ and
$\bar\alpha=1-\alpha$ are the fractions of the light-cone momentum of the
parton $a$ carried by the partons $b$ and $c$ respectively (the notation used
further on). The intuitive motivation for this prescription is rather
transparent.  The incident constituent parton $a$ contains different Fock
components, the bare parton $|a\ra$, the excitation $|bc\ra$, etc. If the
states $|a\ra$ and $|bc\ra$ had the same interaction amplitudes, no
production process $a\to bc$ would be possible, since the composition of the
coherent Fock components which build up the constituent parton $a$ remained
unchanged.  Thus, the {\it production} amplitude of $a\to bc$ is proportional
to the difference of the {\it interaction} amplitudes $f_a-f_{(bc)}$ which
coincides with the scattering amplitude $f_{(\bar abc)}$. Remarkably, each of
the amplitudes $f_a$ and $f_{(bc)}$ might be infra-red divergent (like in the
case of $G\to\bar cc$), while $f_{(\bar abc)}$ is infra-red stable due to
color neutrality of the system.

Treating the process Eq.~(\ref{100}) in more detail, one should
discriminate between three different color and spin states in which the
$\bar cc$ pair can be produced.  These states are orthogonal and do not
interfere in the cross section. They include:
 \begin{enumerate}
 \item The color-singlet $C$-even $\bar cc$ state. The corresponding
amplitude is odd (O) relative to simultaneous permutation of spatial
and spin variables of the $\bar cc$ and has the form,
 \beq
A^{\bar\mu\mu}_{ij,a}(\vec\kappa,\vec k_T,\alpha) =
\sum\limits_{e=1}^8 {1\over6}\,
\delta_{ae}\,\delta_{ij}\ O^{\bar\mu\mu}_{e}
(\vec\kappa,\vec k_T,\alpha)\ .
\label{200}
 \eeq
 Here $\vec\kappa$ and $\vec k_T$ are the relative and total transverse
momenta of the $\bar cc$ pair respectively; $\mu,\bar\mu$ are spin indexes,
$a$ and $i,j$ are color indexes of the gluon and produced quarks,
respectively. We will classify such a state as $1^-$, which means a color
singlet with odd parity relative to index permutation. Note the $1^+$ cannot
be produced in the reaction Eq.~(\ref{100}). 

 \item Color-octet $\bar cc$ state with the production amplitude also
antisymmetric relative simultaneous permutation of spatial and spin variables
of the $\bar cc$ ($8^-$),
 \beq
B^{\bar\mu\mu}_{ij,a}(\vec\kappa,\vec k_T,\alpha) =
\sum\limits_{e,g=1}^8 {1\over2}\,
d_{aeg}\,\tau_{g}(ij)\ O^{\bar\mu\mu}_{e}
(\vec\kappa,\vec k_T,\alpha)\ .
\label{300}
 \eeq
 Here $\lambda_g = \tau_g/2$ are the Gell-Mann matrixes.

 \item Color-octet $\bar cc$ with the amplitude symmetric relative
permutation of quark variables ($8^+$),
 \beq
C^{\bar\mu\mu}_{ij,a}(\vec\kappa,\vec k_T,\alpha) =
\sum\limits_{e,g=1}^8 {i\over2}\,
f_{aeg}\,\tau_{g}(ij)\ E^{\bar\mu\mu}_{e}
(\vec\kappa,\vec k_T,\alpha)\ .
\label{320}
 \eeq
\end{enumerate}
 
The amplitudes Eq.~(\ref{200}) -- (\ref{300}) contain the common 
factor
 \beq
O^{\bar\mu\mu}_{e}(\vec\kappa,\vec k_T,\alpha) = 
\int d^2r\,d^2s\,e^{i\vec\kappa\cdot\vec r -
i\vec k_T\cdot\vec s}\,
\Psi_{\bar cc}^{\bar\mu\mu}(\vec r)\,
\Bigl[\gamma^{(l)}(\vec s-\alpha\vec r) -
\gamma^{(l)}[\vec s+\bar\alpha\vec r]\Bigr]\ ,
\label{340}
 \eeq
which is odd (O) under permutation of the non-color variable of the 
quarks. Correspondingly, the even (E) factor in the amplitude 
Eq.(\ref{320}) reads,
 \beq
E^{\bar\mu\mu}_{e}(\vec\kappa,\vec k_T,\alpha) = 
\int d^2r\,d^2s\,e^{i\vec\kappa\cdot\vec r +
i\vec k_T\cdot\vec s}\,
\Psi_{\bar cc}^{\bar\mu\mu}(\vec r)\,
\Bigl[\gamma^{(l)}(\vec s-\alpha\vec r) +
\gamma^{(l)}[\vec s+\bar\alpha\vec r]
- 2\,\gamma^{(l)}(\vec s)\Bigr]\ ,
\label{360}
 \eeq
 Here $\vec s$ and $\vec r$ is the position of the center of gravity and
the relative transverse separation of the $\bar cc$ pair, respectively.
The LC wave function $\Psi_{\bar cc}^{\bar\mu\mu}(\vec r)$ of the $\bar
cc$ component of the incident gluon in Eqs.~(\ref{340})-(\ref{360}) reads,
 \beq
\Psi_{\bar cc}^{\bar\mu\mu}(\vec r) = 
\frac{\sqrt{2\,\alpha_s}}{4\pi}\,
\xi^\mu\,\hat\Gamma\,\tilde\xi^{\bar\mu}\,
K_0(m_c r)\ ,
\label{370}
 \eeq
 where the vertex operator has the form,
 \beq
\hat\Gamma = m_c\,\vec\sigma\cdot\vec e +
i(1-2\alpha)\,(\vec\sigma\cdot\vec n)
(\vec e\cdot\vec\nabla) +
(\vec n\times\vec e)\cdot\vec\nabla\ ,
\label{370a}
 \eeq
 where $\vec\nabla = d/d\vec r$; $\alpha$ is the fraction of the
gluon light-cone momentum carried by the $c$ quark; $\vec e$ is
the polarization vector of the gluon; and $m_c$ is the $c$-quark mass.

The profile function $\gamma^{(e)}(\vec s)$ in Eqs.~(\ref{340}) 
--(\ref{360}) is related by Fourier transformation to the amplitude
$F^{(e)}(\vec k_T,\{X\})$, of absorption of a real gluon by a nucleon,
$GN\to X$, which also can be treated as an "elastic"  (color-exchange)
gluon-nucleon scattering with momentum transfer $\vec k_T$,
 \beq
\gamma^{(e)}(\vec s) =
\frac{\sqrt{\alpha_s}}{2\pi\sqrt{6}}
\int \frac{d^2k_T}{k_T^2+\lambda^2}\,
e^{-i\vec k_T\cdot\vec s}\,
F^{(e)}_{GN\to X}(\vec k_T,\{X\})\ ,
\label{380}
 \eeq
 where the upper index $(e)$ shows the color polarization of the gluon, and
the variables $\{X\}$ characterize the final state $X$ including the color of
the scattered gluon. In what follows we assume dependence on the variables
$\{X\}$ implicitly. 

It is important for further consideration to relate the profile function
(\ref{380}) to the unintegrated gluon density ${\cal F}(k_T,x)$ and to the
dipole cross section $\sigma_{\bar qq}(r,x)$,
 \beqn
&& \int d^2b\,d\{X\}\,
\sum\limits_{d=1}^8 \left|\gamma^{(e)}(\vec s + \vec r) -  
\gamma^{(e)}(\vec s)\right|^2
\nonumber\\
&=& \frac{4\pi}{3}\,\alpha_s\,
\int \frac{d^2k_T}{k_T^2+\lambda^2}\,
\left(1 - e^{i\vec k_T\cdot\vec r}\right)\,
{\cal F}(k_T,x_2) = \sigma_{\bar qq}(r,x_2)\ ,
\label{390}
 \eeqn
 where $x_2= M_{\bar cc}^2/(2 m_N E_G)$ in the nucleon rest frame.

Let us consider the production cross sections of a $\bar cc$ pair in each
of three states listed above, Eqs.~(\ref{200})--(\ref{320}). The cross
section of a color-singlet $\bar cc$ pair, averaged over polarization and
colors of the incident gluon reads,
 \beq
\sigma^{(1)} = \frac{1}{(2\pi)^4}
\sum\limits_{\mu,\bar\mu,i,j}\ 
\int\limits_0^1 d\alpha\int d^2\kappa\,
d^2k_T\, \overline{\Bigl|
A^{\bar\mu\mu}_{ij,a}(\vec\kappa,\vec k_T,\alpha)
\Bigr|^2}
\label{400}
 \eeq
 Using relations (\ref{2}), (\ref{340}), (\ref{370}) and (\ref{390})
this relation can be modified as,
 \beq
\sigma^{(1)} =
\sum\limits_{\mu,\bar\mu}\,\int\limits_0^1 
d\alpha\int d^2r\, \sigma_1(r,\alpha)\,
\Bigl|\Psi^{\mu\bar\mu}(\vec r,\alpha)
\Bigr|^2\ ,
\label{410}
 \eeq
 where
 \beq
\sigma_1(r,\alpha) = {1\over8}\,\sigma_{\bar qq}(r,x_2)\ ;
\label{420}
 \eeq
 \beq 
\sum\limits_{\mu,\bar\mu}\,
\Bigl|\Psi^{\mu\bar\mu}(\vec r,\alpha)
\Bigr|^2 \,=\,
\frac{\alpha_s}{(2\pi)^2}\,
\Bigl[m_c^2\,K_0^2(m_c r) + 
(\alpha^2 + \bar\alpha^2)\,
m_c^2\,K_1^2(m_c r)\Bigr]\ .
\label{420a}
 \eeq

The cross sections of a color-octet $\bar cc$ pair production either 
in $8^-$ (Odd) or $8^+$ (Even) states has the form,
 \beq
\sigma^{(8)}_{O(E)} = \sum\limits_{\mu,\bar\mu}\,
\int\limits_0^1 d\alpha\int d^2r\,
\sigma^{(8)}_{O(E)}(r,\alpha)\,
\Bigl|\Psi^{\mu\bar\mu}(\vec r,\alpha)\Bigr|^2\ ,
\label{430}
 \eeq
 where
 \beqn
\sigma^{(8)}_{O}(r,\alpha,x_2) &=& 
{5\over16}\,\sigma_{\bar qq}^N(r,x_2)\ ;
\label{440}\\
\sigma^{(8)}_{E}(r,\alpha,x_2) &=&
{9\over16}\,\Bigl[ 2\sig(\alpha r,x_2) +
2\sig(\bar\alpha r,x_2) - \sig(r,x_2)\Bigr]\ .
\label{450}
 \eeqn

The total cross section of a $\bar cc$ pair production reads,
 \beq
\sigma(GN\to\bar ccX)\equiv \sigma^{(1)} + 
\sigma^{(8)}_O + \sigma^{(8)}_E = \sum\limits_{\mu,\bar\mu}\,
\int\limits_0^1 d\alpha\int d^2r\,
\sigma_3(r,\alpha,x_2)\,
\Bigl|\Psi^{\mu\bar\mu}(\vec r,\alpha)\Bigr|^2\ ,
\label{460}
 \eeq
 where
 \beq
\sigma_3(r,\alpha,x_2) = {9\over8}\,\Bigl[
\sig(\alpha r,x_2)+\sig(\bar\alpha r,x_2)\Bigr] -
{1\over8}\,\sig(r,x_2)\ .
\label{470}
 \eeq 

In order to estimate the relative yield of the $1^-$, $8^-$ and $8^+$
states we can rely upon the approximation $\sig(r)\propto r^2$ which is
rather accurate in the case of a $\bar cc$ pair, since its separation
$r\sim1/m_c$ is small. Then we derive,
 \beq
\sigma^{(1)}\ :\ \sigma^{(8)}_O\ :\ \sigma^{(8)}_E =
1\ :\ {5\over2}\ :\ {117\over70}\ .
\label{480}
 \eeq
 Thus, about $20\%$ of the produced $\bar cc$ pairs are in a color-singlet
state, the rest are color-octets. One can calculate the inclusive cross
section of open charm production in pp collision multiplying the
color-octet part of the cross section Eq.~(\ref{460}) by the gluon
distribution in the proton,
 \beqn
\frac{d\sigma(pp\to\{\bar cc\}_8 X)}{d\,y} &=& 
{9\over8}\,G(x_1)\,\sum\limits_{\mu,\bar\mu}\,
\int\limits_0^1 d\alpha\int d^2r\,
\Bigl|\Psi^{\mu\bar\mu}(\vec r,\alpha)\Bigr|^2
\nonumber\\ &\times&
\left[\sig(\alpha r,x_2)+\sig(\bar\alpha r,x_2)-
{1\over4}\,\sig(r,x_2)\right]\ ,
 \label{485}
 \eeqn
 where $G(x_1)=x_1\,g(x_1)$ and
 \beq
x_1=\frac{M_{\bar cc}^2}{x_2\,s}\ .
\label{485a}
 \eeq
 Note that not all the color-singlet pairs end up with charmonium states,
but some may decay to open charm channels and end up with $D\bar D$. 
We come back to this problem in Sect.~\ref{singlet}.

\subsection{Multiple color-exchange interactions and production of a
\boldmath$\bar cc$ pair in nuclear matter}\label{eikonal}

An important advantage of the LC dipole approach is the simplicity of
calculations of nuclear effects. Since partonic dipoles are the eigenstates
of interaction one can simply eikonalize the cross section on a nucleon
target \cite{zkl} provided that the dipole size is ``frozen'' by Lorentz time
dilation. Therefore, the cross section of $\bar cc$ pair production off a
nucleus has the form \cite{npz},
 \beq
\sigma(GA\to \bar ccX) = 
2\,\sum\limits_{\mu,\bar\mu}\,\int d^2b\int d^2r
\int\limits_0^1 d\alpha\,
\Bigl|\Psi^{\mu\bar\mu}(\vec r,\alpha)\Bigr|^2\,
\left\{1\,-\,\exp\left[-{1\over2}\,
\sigma_3(r,\alpha,x_2)\,T_A(b)\right]\right\}\ .
\label{500}
 \eeq
 Apparently, this expression leads to shadowing correction which is a
higher twist effect and vanishes as $1/m_c^2$.  Indeed, it was found in
\cite{npz} that in the kinematic range of fixed target experiments at the
Tevatron, Fermilab, $x_2\sim 10^{-2},\ x_F\sim0.5$, the shadowing effects
are rather weak even for heavy nuclei,
 \beq
1-R_A \lsim 0.05\ ,
\label{510}
 \eeq
 where $R_A$ is defined in (\ref{2}).

On the other hand, a substantial shadowing effect, several times stronger
than (\ref{510}) was found in \cite{kth} for charmonium production,
although it is also a higher twist effect.  In the case of open charm
production there are additional cancellations which grossly diminish
shadowing. The smallness of the effect maybe considered as a justification
for the parton model prescription to neglect this correction as a
higher twist effect. However, 
the dipole cross section $\sig(r,x_2)$ steeply
rises with $1/x_2$ especially at small $r$ and and the shadowing
corrections increase reaching values of about $10\%$ at $x_2=10^{-3}$, and
about $30\%$ at $x_2=10^{-5}$. 

\section{Gluon shadowing}\label{shadowing}

The phenomenological dipole cross section which enters the exponent in
Eq.~(\ref{500}) is fitted to DIS data. Therefore it includes effects of gluon
radiation which are in fact the source of rising energy ($1/x$) dependence of
the $\sig(r,x)$. However, a simple eikonalization in Eq.~(\ref{500}) 
corresponds to the Bethe-Heitler approximation assuming that the whole
spectrum of gluons is radiated in each interaction independently of other
rescatterings. This is why the higher order terms in expansion of (\ref{500})
contain powers of the dipole cross section.  However, gluons radiated due to
interaction with different bound nucleons can interfere leading to damping of
gluon radiation similar to the Landau-Pomeranchuk \cite{lp} effect in QED. 
Therefore, the eikonal expression Eq.~(\ref{500}) needs corrections which are
known as gluon shadowing.

Nuclear shadowing of gluons is treated by the parton model in the infinite
momentum frame of the nucleus as a result of glue-glue fusion. On the other
hand, in the nuclear rest frame the same phenomenon is expressed in terms of
the Glauber like shadowing for the process of gluon radiation \cite{mueller}. 
In impact parameter representation one can easily sum up all the multiple
scattering corrections which have the simple eikonal form \cite{zkl}.
Besides, one can employ the well developed color dipole phenomenology with
parameters fixed by data from DIS. Gluon shadowing was calculated employing
the light-cone dipole approach for DIS \cite{kst2} and production of
charmonia \cite{kth}, and a substantial deviation from QCD factorization
was found. Here we calculate gluon shadowing for $\bar cc$ pair production.

\subsection{Associated gluon radiation, \boldmath$G\,N \to \bar
cc\,G\,X$}\label{gl-rad}

First of all, one should develop a dipole approach for gluon radiation
accompanying production of a $\bar cc$ pair in gluon-nucleon collision. 
Then nuclear effects can be easily calculated via simple eikonalization.

\begin{figure}[tbh]
\includegraphics{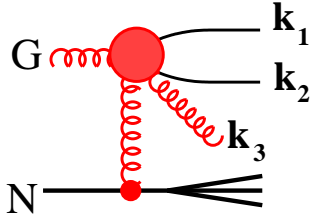} 
\begin{center} 
\vspace{3.7cm} 
\parbox{13cm}
 {\caption[Delta]
 {\sl Perturbative QCD mechanism for production of a $\bar cc$ pair and a
gluon in a gluon-nucleon collision. The upper blob includes different
attachments of the gluons as is depicted in Fig.~\ref{graphs}.} 
 \label{1graph}}
\end{center}
 \end{figure}

The amplitude of the process $G\,N \to \bar cc\,G$ is illustrated in
Fig.~\ref{1graph}.
  According to the general prescription \cite{hir} the dipole cross section
which
enters the factorized formula for the process of parton $a$-nucleon
collision leading to multiparton production, $a\,N \to b+c+\dots+d\ X$, is
the
cross section for the colorless multiparton ensemble $|\bar abc\dots
d\ra$. The same multiparton dipole cross section is responsible for
nuclear shadowing. Indeed, in the case of the process $G\,N\to\bar cc\,X$
it was the cross section $\sigma_3$ Eq.~(\ref{470}) which correspond to a
state $|\bar ccG\ra$ interacting with a nucleon. 

Correspondingly, in the case of additional gluon production, $G\to\bar
cc\,G$, it is a 4-parton, $|\bar ccGG\ra$, cross section $\sigma_4(\vec
r,\vec\rho,\alpha_1,\alpha_2,\alpha_3)$. Here $\vec r$ and $\vec\rho$ are the
transverse $\bar cc$ separation and the distance between the $\bar cc$ center
of gravity and the final gluon, respectively. Correspondingly,
$\alpha_1=\alpha_c$, $\alpha_2=\alpha_{\bar c}$, and $\alpha_3=\alpha_G$.

Treating the charm quark mass as a large scale, one can neglect
$r\ll\rho$, then the complicated expression for $\sigma_4$ becomes rather
simple. One should disentangle two possibilities. 

\begin{enumerate}

\item The $\bar cc$ pair is in a color-singlet state. At $r=0$ it does not
participate in the interaction with the nucleon target, therefore in this 
limit,
 \beq
\sigma_4^{(1)} = \sigma_{GG}\Bigl(\bar\alpha_3\rho\Bigr)\ .
\label{520}
 \eeq
 Here $\sigma_{GG}$ is the cross section of interaction of a glue-glue
dipole with a nucleon. Although we do not rely on perturbative methods of
calculation of the dipole cross section, we will assume that the relation
between the $\bar qq$ and $GG$ dipole cross sections is given by the
simple Casimir factor,
 \beq
\sigma_{GG}(r)={9\over4}\,\sig(r)\ .
\label{530}
 \eeq

\item
The $\bar cc$ pair is in an octet-color state, then at $r=0$ it is
indistinguishable from a gluon. Therefore, the 4-parton cross section
becomes a 3-gluon one which has the form,
 \beq
\sigma_4^{(8)} = 
{1\over2}\,\left[\sigma_{GG}(\rho) +
\sigma_{GG}(\alpha_3\rho) + 
\sigma_{GG}\Bigl(\bar\alpha_3\rho\Bigr)\right]\ .
\label{540}
 \eeq
\end{enumerate}

Remarkably, in the limit $\alpha_3\ll1$ we are interested in,
 \beq
\sigma_4^{(1)}\Bigr|_{\alpha_3\to0} = 
\sigma_4^{(8)}\Bigr|_{\alpha_3\to0} =
\sigma_{GG}(\rho)\ ,
\label{550}
 \eeq
 i.e. the cross section is independent of the color state of the $\bar cc$
pair in this limit.

As far as the cross section $\sigma_4$ and the $\bar ccG$ LC wave function,
$\Psi^{\bar\mu\mu}_{ij,b(a)}(\vec r,\vec\rho,\alpha_c,\alpha_G,\vec e_f)$,
are known, the cross section of the reaction $G_aN\to \bar ccG_bX$ takes the
form,
 \beqn 
\sigma(G_aN\to \bar ccG_b) &=& 
\sum\limits_{\vec e_f} \sum\limits_{\bar\mu,\mu=1}^2 
\sum\limits_{i,j=1}^{N_c} \sum\limits_{b=1}^{N_c^2-1} 
\,\int d^2r\,d^2\rho\,d\alpha_c\,d\alpha_G
\nonumber\\ &\times& 
\Bigl|\Psi^{\bar\mu\mu}_{ij,b(a)}
(\vec r,\vec\rho,\alpha_c,\alpha_G,\vec e_f)
\Bigr|^2\, \sigma_4(\vec r,\vec\rho,\alpha_c,\alpha_G)\ .
\label{570}
 \eeqn

 The analysis of the structure of the amplitude $G_,N\,\to\,\bar
cc\,G_b\,X$ is performed in the Appendix~A in the leading order in
$\alpha_s$ assuming $\alpha_G\ll\lambda^2/M_{\bar cc}^2\ll1$ and $r\sim
m_c \ll 1/\Lambda_{QCD}$. The amplitude Eq.~(\ref{a230}) contains in the
curly brackets three terms which correspond to three different states
of the $\bar cc$ pair and to the following LC wave functions of the $\bar
ccG$ system. 

a) The $\bar cc$ pair is in a color-singlet asymmetric state $(1^-)$.
 \beq
\Psi^{\bar\mu\mu}_{ij,b(a)}
(\vec r,\vec\rho,\alpha_c,\alpha_G,\vec e_f) =
\frac{1}{\sqrt{3}}\,\delta_{ab}\,\delta_{ij}\,
\Psi^{\bar\mu\mu}(\vec r,\alpha_c)\,
\Bigl[\Phi^{(1^-)}_{cG}(\vec\rho+\alpha_c\vec r) -
\Phi^{(1^-)}_{cG}(\vec\rho-\bar\alpha_c\vec r)\Bigr]\ .
\label{590}
 \eeq

b) The $\bar cc$ pair is in a color-octet asymmetric state $(8^-)$.
 \beq
\Psi^{\bar\mu\mu}_{ij,b(a)}
(\vec r,\vec\rho,\alpha_c,\alpha_G,\vec e_f) =
\sqrt{3}\sum\limits_{g=1}^{N_c^2-1} 
d_{abg}(\tau_g)_{ij}\,
\Psi^{\bar\mu\mu}(\vec r,\alpha_c)\,
\Bigl[\Phi^{(8^-)}_{cG}(\vec\rho+\alpha_c\vec r) -
\Phi^{(8^-)}_{cG}(\vec\rho-\bar\alpha_c\vec r)\Bigr]\ .
\label{610}
 \eeq

c) The $\bar cc$ pair is in a color-octet symmetric state $(8^+)$.
 \beqn
\Psi^{\bar\mu\mu}_{ij,b(a)}
(\vec r,\vec\rho,\alpha_c,\alpha_G,\vec e_f) &=&
i\sqrt{3}\sum\limits_{g=1}^{N_c^2-1} 
f_{abg}(\tau_g)_{ij}\,
\Psi^{\bar\mu\mu}(\vec r,\alpha_c)\nonumber\\&\times&
\Bigl[\Phi^{(8^+)}_{cG}(\vec\rho+\alpha_c\vec r) +
\Phi^{(8^+)}_{cG}(\vec\rho-\bar\alpha_c\vec r)
-2\,\Phi^{(8^+)}_{cG}(\vec\rho)\Bigr]\ .
\label{630}
 \eeqn

 Within perturbative QCD used so far and in Appendix~A, the functions
$\Phi_{cG}(\vec\rho)$ are equal for different states, $1^-$, $8^-$ and
$8^+$. 
 \beq
\Phi^{(1^-)}_{cG}(\vec\rho)=
\Phi^{(8^-)}_{cG}(\vec\rho)=
\Phi^{(8^+)}_{cG}(\vec\rho)=
\frac{i\sqrt{\alpha_s}}{\sqrt{3}\,\pi}\
\vec e_f\cdot\vec\nabla\ 
K_0(\tau\rho)\ ,
\label{650}
 \eeq
where
 \beq
\tau^2=\lambda^2+\alpha_G\,M_{\bar cc}^2\ .
\label{670}
 \eeq
 However, the nonperturbative effects considered further in
Sect.~\ref{nonpqcd} make these functions different, so we mark them in
accordance with the color state of the $\bar cc$. Note that
$\Phi_{cG}(\vec\rho)$ is different from the LC wave function of a quark-gluon
system, this is why $\tau$ depends on $M_{\bar cc}$. 

Relying on smallness of $r\sim 1/m_c\ll 1/\tau$ the combinations of the
functions $\Phi_{cG}$ in (\ref{590}) - (\ref{630}) can be simplified,
 \beq
\Phi_{cG}(\vec\rho+\alpha_c\vec r) -
\Phi_{cG}(\vec\rho-\bar\alpha_c\vec r)\approx
\vec r\cdot\vec\nabla\ \Phi_{cG}(\vec\rho)\ ;
\label{690}
 \eeq 
 \beq
\Phi_{cG}(\vec\rho+\alpha_c\vec r) +
\Phi_{cG}(\vec\rho-\bar\alpha_c\vec r)\approx
- 2\Phi_{cG}(\vec\rho) \approx
(2\alpha_c-1)\,\vec r\cdot\vec\nabla\ \Phi_{cG}(\vec\rho)\ .
\label{690a}
 \eeq
 We use this approximation in what follows.

\subsection{Reaction \boldmath$G\,A\to \bar cc\,G\,X$ and gluon density
in nuclei}\label{gl-rad-nucl}

Precess of gluon radiation off nuclei is related to gluon shadowing. It was
first suggested in \cite{mueller} to use a process $j\,A\to GG\,X$, where $j$
is a colorless current, as a probe for gluon distribution in the nucleus.
There are more hard reactions with gluon radiation which can be used as a
probe for gluon shadowing (see e.g. in \cite{kst2,kth}).  However, one should
be cautious relying on the process independence of gluon shadowing which
follows from QCD factorization.  It is not obvious that the charm mass scale
is sufficiently high to neglect higher twist effects. It is especially
problematic for gluon radiation which involves a semi-hard scale
\cite{kst2}.  Indeed, it is demonstrated in \cite{kth} that gluon shadowing
in charmonium production is much stronger than in DIS \cite{kst2}. 
Therefore, gluon shadowing should be calculated separately for each color
state of the produced $\bar cc$ pair.

Following \cite{krt1,kst2,krt2,knst,krtj} the cross section of the
reaction $G\,A\to \bar cc\,G\,X$ is presented in the form,
 \beq
\sigma(GA\to \bar ccGX) = 
A\,\sigma(GN\to \bar ccGX) - \Delta\sigma_G\ ,
\label{800}
 \eeq
 where  the second term $\Delta\sigma_G$ describing shadowing of gluon 
density has the form,
 \beqn
&& \Delta\sigma_G = {1\over2}\,{\rm Re} \int
d^2r_1\,d^2r_2\,d^2\rho_1\,d^2\rho_2\,d^2b\,
dz_1\,dz_2\,d\alpha_c\,d(\ln\alpha_G)\nonumber\\
&\times&\sum\limits_{\vec e_f}\sum\limits_{\mu,\bar\mu=1}^{2}
\sum\limits_{i,j=1}^{N_c}
\sum\limits_{b=1}^{N-c^2-1}
\Psi^{{\bar\mu\mu}^\dagger}_{ij,b(a)}
(\vec r_2,\vec\rho_2,\alpha_c,\alpha_G,\vec e_f)\,
\sigma_4(\vec r_2,\vec\rho_2,\alpha_c,\alpha_G)\,
G_{\bar ccG}(\vec r_2,\vec\rho_2,z_2;\vec r_1,\vec\rho_1,z_1)
\nonumber\\ &\times&
\sigma_4(\vec r_1,\vec\rho_1,\alpha_c,\alpha_G)\,
\Psi^{{\bar\mu\mu}^\dagger}_{ij,b(a)}
(\vec r_1,\vec\rho_1,\alpha_c,\alpha_G,\vec e_f)\,
\rho_A(b,z_2)\,\rho_A(b,z_1)\,
\Theta(z_2-z_1)\ .
\label{820}
 \eeqn
 Here $G_{\bar ccG}(\vec r_2,\vec\rho_2,z_2;\vec r_1,\vec\rho_1,z_1)$ is
the LC Green function describing evolution of the $\bar ccG$
system from initial separations $\vec r_1,\ \vec\rho_1$ at the point
$z_1$ to final $\vec r_2,\vec\rho_2$ at the point $z_2$. It obeys the
Schr\"odinger type equation,
 \beqn
i\,\frac{\partial}{\partial z_2}\,
G_{\bar ccG}(\vec r_2,\vec\rho_2,z_2;\vec r_1,\vec\rho_1,z_1) &=&
\left[ -\frac{\Delta_{r_2}}{2\nu\alpha_c\bar\alpha_c}
-\frac{\Delta_{\rho_2}}{2\nu\alpha_G} -
{i\over2}\,\sigma_4(\vec r_2,\vec\rho_2,\alpha_c,\alpha_G)\,
\rho_A(b,z_2)\right]\nonumber\\ &\times&
G_{\bar ccG}(\vec r_2,\vec\rho_2,z_2;\vec r_1,\vec\rho_1,z_1)\ ,
\label{840}
 \eeqn
 where $\nu$ is the energy of the incident gluon, and the boundary 
condition is
 \beq
G_{\bar ccG}(\vec r_2,\vec\rho_2,z_2;\vec r_1,\vec\rho_1,z_1)
\Bigr|_{z_2=z_1} = \delta(\vec r_2-\vec r_1)\,
\delta(\vec\rho_2-\vec\rho_1)\ .
\label{860}
 \eeq

The dominant contribution to integration in $\alpha_G$ in (\ref{820})
comes from the region of small $\alpha_G \lsim \lambda^2/M_{\bar cc}^2
\ll 1$, therefore, according to (\ref{550}) we can assume in (\ref{820})  
and (\ref{840}) that
 \beq
\sigma_4(\vec r,\vec\rho,\alpha_c,\alpha_G) =
\sigma_{GG}(\rho)\ .
\label{880}
 \eeq
 In this case the Green function factorizes as
 \beq
G_{\bar ccG}(\vec r_2,\vec\rho_2,z_2;\vec r_1,\vec\rho_1,z_1)=
G_{\bar cc}(\vec r_2,z_2;\vec r_1,z_1)\,
G_{(\bar cc)G}(\vec\rho_2,z_2;\vec\rho_1,z_1)\ ,
\label{900}
 \eeq
 where $G_{\bar cc}$  describes the intrinsic life
of the $\bar cc$ pair, while $G_{(\bar cc)G}$ controls the relative 
motion of the point-like $\bar cc$ pair (when its size is neglected) and the 
gluon. These two-body Green functions obey the following equations,
 \beq
i\,\frac{\partial}{\partial z_2}\,
G_{\bar cc}(\vec r_2,z_2;\vec r_1,z_1) = \left[
-\frac{\Delta_{\vec r_2}}{2\nu\alpha_c\bar\alpha_c}\ 
+V_{\bar cc}(r_2,z_2)\right]
G_{\bar cc}(\vec r_2,z_2;\vec r_1,z_1)\ ,
\label{920}
\eeq
\beq
i\,\frac{\partial}{\partial z_2}\, 
G_{(\bar cc)G}(\vec\rho_2,z_2;\vec\rho_1,z_1) =
\left[
-\frac{\Delta_{\vec\rho2}}{2\nu\alpha_G\bar\alpha_G}
+V_{GG}(\rho_2,z_2)\right]
G_{(\bar cc)G}(\vec\rho_2,z_2;\vec\rho_1,z_1)\ ,
\label{940}
 \eeq
with boundary conditions
 \beqn
G_{\bar cc}(\vec r_2,z_2;\vec r_1,z_1)\Bigr|_{z_2=z_1}
&=& \delta(\vec r_2-\vec r_1)\ ,\nonumber\\
G_{(\bar cc)G}(\vec\rho_2,z_2;\vec\rho_1,z_1)\Bigr|_{z_2=z_1}
&=& \delta(\vec\rho_2-\vec\rho_1)\ .
\label{960}
 \eeqn

The imaginary parts of the light-cone potentials in (\ref{920}) and
(\ref{940})  describe absorption of a $\bar cc$ or $GG$ pairs propagating
through the nucleus. 
 \beqn
{\rm Im}\,V_{\bar cc}(r_2,z_2) &=& 
-{1\over2}\,\sigma_{\bar qq}(r_2)\,\rho_A(b,z_2)\ ;
\label{970}\\
{\rm Im}\,V_{\bar cc}(\rho_2,z_2) &=& 
-{1\over2}\,\sigma_{GG}(\rho_2)\,\rho_A(b,z_2)\ ;
\label{975}
 \eeqn

 The real parts of the potentials take into account the interaction between
the partons and in particular incorporate phenomenologically confinement. 
This interaction is very important for massless gluons and is considered in
next section. At the same time, for a heavy $\bar cc$ both the interaction
potential and absorption cross section are small and can be neglected. Then,
the Green function $G_{\bar cc}(\vec r_2,z_2;\vec r_1,z_1)$ describes the
free propagation of a $\bar cc$ pair, and the solution of Eq.~(\ref{920}) has
the analytical form \cite{feynman},
 \beq
G_{\bar cc}(\vec r_2,z_2;\vec r_1,z_1) =
\frac{\nu\alpha_c\bar\alpha_c}{2i\pi(z_2-z_1)}\ 
\exp\left[-\frac{\nu\alpha_c\bar\alpha_c
(\vec r_2-\vec r_1)^2}{2i(z_2-z_1)}\right]\ .
\label{980}
 \eeq
 In the high-energy limit, $\nu\to\infty$, all fluctuations of the size 
of the $\bar cc$ pair are ``frozen'' by Lorentz time dilation for the 
time of propagation through the nucleus, therefore,
 \beq
G_{\bar cc}(\vec r_2,z_2;\vec r_1,z_1)\Bigr|_{\nu\to\infty} 
= \delta(\vec r_2-\vec r_1)\ .
\label{1000}
 \eeq
 In this limit also the Green function $G_{(\bar
cc)G}(\vec\rho_2,z_2;\vec\rho_1,z_1)$ which describes relative motion of
the point-like color-octet $\bar cc$ and the gluon takes the simple form,
 \beq
G_{(\bar cc)G}(\vec\rho_2,z_2;\vec\rho_1,z_1)\Bigr|_{\nu\to\infty}
= \delta(\vec\rho_2-\vec\rho_1)\,
\exp\left[-{1\over2}\,\sigma_{GG}(\rho_2)
\int\limits_{z_1}^{z_2} dz\,\rho_A(b,z)\right]\ .
\label{1020}
 \eeq

Now we can switch back to the gluon shadowing correction $\Delta\sigma_G$
in Eq.~(\ref{800}) related to coherence effects in gluon radiation by a
$\bar cc$ pair propagating through the nucleus. The result may depend on
the state in which the $\bar cc$ pair is left after the gluon is radiated.
We
classified above those states as $1^-$, $8^-$ and $8^+$. In what follows
we mark these states by index $k=1^-,8^-$ or $8^+$. In the limit
$\nu\to\infty$ the corresponding gluon shadowing corrections to the cross
section of the process $G\,A \to \bar cc\,G X$ in Eq.~(\ref{800}) read,
 \beq
\Delta\sigma^{(k)}_G = S^{(k)}\,\sigma(GN\to \bar ccGX)\ ,
\label{1040}
 \eeq
 where
 \beqn
 S^{(k)} &=& \frac{1}{4\,C} \int 
d^2b\,d^2\rho_2\,d^2\rho_1\,dz_2\,dz_1\,
\vec\nabla \Phi^{(k)}_{cG}(\vec\rho_2,\alpha_G/\alpha_c)\,
\sigma_{GG}(\vec\rho_2)\,
G_{(\bar cc)G}(\vec\rho_2,z_2;\vec\rho_1,z_1)\,
\sigma_{GG}(\vec\rho_1)\nonumber\\ &\times&
\vec\nabla \Phi^{(k)}_{cG}(\vec\rho_1,\alpha_G/\alpha_c)\,
\rho_A(b,z_2)\,\rho_A(b,z_1)\,\Theta(z_2-z_1)\ .
\label{1060}
 \eeqn
 Thus, the entire dependence on the state "$k$" in which the $\bar cc$ pair
is produced, comes via the wave functions $\Phi^{(k)}_{cG}$ which varies with
$k$ only if the nonperturbative interaction between partons matters and is
sensitive to $k$. We discuss the influence of this nonperturbative effects in
more detail below. 

At finite energies $\nu$ the size of the $\bar cc$ system can fluctuate
during propagation through the nucleus.  In order to incorporate this effect
one should introduce under the integral in Eq.~(\ref{1060})  an extra factor
 \beq
I^{(k)} = \frac{N^{(k)}}{D^{(k)}}\ ,
\label{1080}
 \eeq
where
 \beq
N^{(k)} = \sum\limits_{\bar\mu\mu} 
\int d^2 r_1\,d^2r_2\,d\alpha_c\,
\Psi^{{\bar\mu\mu}^\dagger}(\vec r_2,\alpha_c)\,
G_{\bar cc}(\vec r_2,z_2;\vec r_1,z_1)\,
\Psi^{\bar\mu\mu}(\vec r_1,\alpha_c)\,
(\vec r_1\cdot\vec r_2)\, f^{(k)}(\alpha_c)\ ,
\label{1100}
 \eeq
 \beq
D^{(k)} = \sum\limits_{\bar\mu\mu}
\int d^2 r\,d\alpha_c\,
\left|\Psi^{\bar\mu\mu}(\vec r,\alpha_c)\right|^2\,
r^2\,f^{(k)}(\alpha_c)\ .
\label{1120}
 \eeq
 Here the $k$-dependent factor $f^{(k)}$ reads, 
\beq 
f^{(k)}(\alpha_c) =
\left\{ 
\begin{array}{cc} 
1&\ \ \ \ \ \ \ {\rm for\ the\ states}\ 1^-,\ 8^- \\
(2\alpha_c-1)^2&\ \ \,{\rm for\ the\ state}\ \ 8^+\ . 
\end{array}
\right.
\label{1140} 
 \eeq
 Apparently, $I^{(k)}\to 1$ at $\nu\to\infty$.

Using the analytical expressions Eq.~(\ref{980}) for $G_{\bar cc}$ and
Eq.~(\ref{370}) for $\Psi^{\bar\mu\mu}$ we arrive at
 \beq
I^{(k)} = \frac{\int_{\beta}^{1-\beta}
d\alpha_c\,f^{(k)}(\alpha_c)\,
\{E_2(w) + 2\alpha_c\bar\alpha_c[2E_3(w)-2E_4(w)-
E_2(w)]\}}
{\int_0^1 d\alpha_c\,f^{(k)}(\alpha_c)\,
[1 - 4\alpha_c\bar\alpha_c/3]}\ ,
\label{1160}
 \eeq
 where $E_n(w)$ are the integral exponential functions,
 \beq
E_n(w) = \int\limits_1^\infty \frac{dt}{t^n}\,
e^{-wt}\ ,
\label{1180}
 \eeq
 \beqn
w &=&\frac{im_c^2(z_2-z_1)}
{2\nu\alpha_c\bar\alpha_c}\nonumber\\
\beta &=& \frac{m_c}{2\nu}\ .
\label{2000}
 \eeqn
 A remarkable observation is that for finite formation times for the $\bar
ccG$ system the gluon shadowing factor $S^{(k)}$ varies with $k$ even in
perturbative case when the partons propagate as free particles and the
function $\Phi^{(k)}$ is independent of $k$. In the limit of $\nu\to\infty$
the universality of perturbative gluon shadowing is restored.

 \subsection{Self-interacting \boldmath$\bar ccG$}\label{nonpqcd}

One can treat partons as free only if their transverse momenta are
sufficiently large, otherwise the nonperturbative interaction between partons
may generate power corrections \cite{kst2}. Apparently, the softer the
process is, the more important are these corrections. In particular,
diffraction and nuclear shadowing are very sensitive to these effects.
Indeed, the cross section of diffractive dissociation to large masses (so
called triple-Pomeron contribution)  is proportional to the fourth power of
the size of the partonic fluctuation.  Therefore, the attractive
nonperturbative interaction between the partons squeezes the fluctuation and
can substantially reduce the diffractive cross section. Smallness of the
transverse separation in the quark-gluon fluctuation is the only known
explanation for the observed suppression of the diffractive cross section,
which is also known as the problem of smallness of the triple-Pomeron
coupling.  While no data sensitive to gluon shadowing are available yet, a
vast amount of high accuracy diffraction data can be used to fix the
parameters of the nonperturbative interaction.

To solve the light-cone Schr\"odinger equation for the Green function
analytically we use the oscillator form of the light-cone potential of
interaction between the gluon (with small $\alpha_G\ll1$) and quark,
 \beq
\Re V_{GG}(\rho) = \frac{b_0^4\,\rho^2}
{2\,\nu\,\alpha_G\bar\alpha_G}\ .
\label{2020}
 \eeq
With such a form of the potential one can get an analytical solution
of Eq.~(\ref{940}) assuming also that $\sigma_{GG}(r,s)\approx
C_{GG}(s)\,r^2$, where $C_{GG}(s) = d\,\sigma_{GG}(r,s)/d\,r^2_{r=0}$.
The result has a form \cite{kst2},
 \beqn
&& G_{GG}(\vec\rho_2,z_2;\vec \rho_1,z_1) =
\frac{A}{2\pi\,{\rm \sinh}(\Omega\,\Delta z)}
\nonumber\\ &\times&
{\rm exp}\left\{-\frac{A}{2}\,
\left[(\rho_1^2+\rho_2^2)\,{\rm coth}(\Omega\,\Delta z) -
\frac{2\vec\rho_1\cdot\vec\rho_2}{{\rm sinh}(\Omega\,\Delta z)}
\right]\right\}\ ,
\label{2025}
 \eeqn
where
 \beqn
A&=&\sqrt{{b_0}^4-i\,\alpha_G\bar\alpha_G\,\nu\,C_{GG}\,
\rho_A}\nonumber\\
\Omega &=& \frac{i\,A}
{\alpha_G\bar\alpha_G\,\nu}\nonumber\\
\Delta z&=&z_2-z_1\ .
\label{2030}
 \eeqn

With this Green function one also gets the nonperturbative wave function
for
the $qG$ Fock component of the quark \cite{kst2},
 \beq
\Phi_{qG}(\vec\rho) = 
\frac{2\,i}{\pi}\,
\sqrt{\frac{\alpha_s}{3}}\,
\frac{\vec e_f\cdot\vec\rho}{\rho^2}\ 
e^{-b_0^2\rho^2}\ .
\label{2040}
 \eeq
 Fitting the only free parameter $b_0$ to data for large mass single
diffraction $pp\to pX$ which corresponds to diffractive gluon radiation, one
finds a rather strong interaction for gluons, $b_0=0.65\GeV$ \cite{kst2}.
Thus, the mean size of the qluon cloud of a quark turns out to be
surprisingly small, $r_0 = \sqrt{\la\rho^2\ra} = 0.3\fm$.  Such a picture of
the proton which consists of valence quarks surrounded by small size gluon
clouds correctly reproduces the observed energy dependence for total hadronic
cross sections \cite{k3p}. Besides, it goes along with the results of
nonperturbative approaches.  The same size emerges from lattice calculations
for the glue-glue correlation radius \cite{pisa}, from the instanton liquid
model \cite{shuryak} and from the QCD sum rule analysis of the gluonic form
factor of the proton \cite{andreas}.

These results are relevant to and were used in \cite{kst2} for calculation of
gluon shadowing in DIS where a colorless virtual photon fluctuates into the
color-octet-octet system, $\gamma^* \to (\bar qq)_8\,G$.  The nonperturbative
interaction between the partons within a color-octet system $\bar ccG$ which
is under consideration, may be different. Indeed, if the small size, $r\sim
1/m_c$, $\bar cc$ pair is a color singlet, the interaction between the $\bar
cc$ and the gluon should be very weak. In this case one should use the
perturbative expression Eq.~(\ref{650}) for the $\Phi^{(1^-)}_{cG}$ which
corresponds to a large size $\sim 1/\Lambda_{QCD}$ fluctuation and strong
gluon shadowing.  Indeed, it explains the strong nuclear suppression of
$J/\Psi$ production observed at small $x_2$ \cite{kth}.

If, however, the $\bar cc$ pair is a color octet, $8^-$ or $8^+$, and its
size is neglected, then it is equivalent to a gluon and interacts in the same
way. Thus, we came to the problem of interaction between two gluons in one of
the two possible color octet states which have different symmetries relative
permutation of the gluon's color indexes [compare with (\ref{610}) -
(\ref{630})],
 \beqn
|8_a\ra_{GG}^S &=&
\sqrt{{3\over5}}\,\sum\limits_{b,g=1}^{8}
d_{abg}\,|G_b\ra\,|\widetilde G_g\ra\ ;
\label{2060}\\
|8_a\ra_{GG}^A &=&
\frac{1}{\sqrt{3}}\,\sum\limits_{b,g=1}^{8}
f_{abg}\,|G_b\ra\,|\widetilde G_g\ra\ .
\label{2080}
 \eeqn
 The states in the left-hand-side of these equations are the two-gluon Fock
components of a gluon, which are symmetric ($S$), or asymmetric ($A$) relative
permutation of the indexes $b$ and $g$. They correspond to $8^-$ and $8^+$ states
respectively. One of the gluon is marked with a tilde in order to emphasize that
this is a color-octet point-like $\bar cc$ pair, rather than a gluon. 
Correspondingly, the color-singlet state of two gluons which occurs in DIS has the
form,
 \beq
|1\ra_{GG} =
\frac{1}{\sqrt{8}}\,\sum\limits_{b,g=1}^{8}
\delta_{bg}\,|G_b\ra\,|\widetilde G_g\ra\ .
\label{2100}
 \eeq

\begin{figure}[tbh]
\includegraphics{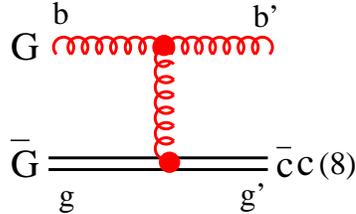}
\begin{center}
\vspace{3cm}
\parbox{13cm}
 {\caption[Delta]
 {\sl Born graph for scattering of a gluon with color 
indexes $b,\ b'$ on a $\bar cc$ pair in color-octet state with indexes 
$g,\ g'$.}
 \label{gg-scatt}}
\end{center}
 \end{figure}

The strength and sign (attractive or repulsive) of the nonperturbative
interactions between the gluons depends on their configuration,
Eqs.~(\ref{2060}) or (\ref{2100}).  to. The potential is known from data
analysis only for the color singlet case Eq.~(\ref{2100}), and it is
impossible to predict the interaction potential for other states having no
knowledge of the interaction dynamics. To establish relations between the
potentials corresponding to the states Eqs.~(\ref{2060}) - (\ref{2100}) we
rely on the relations between ``Coulomb'' perturbative potentials. The
corresponding potential has the form,
 \beq
\hat V_{bg,b'g'}(r) = -\frac{\alpha_s(r)}{r}\,
\sum\limits_{e=1}^8\,f_{beb'}\,f_{geg'}\ .
\label{2120}
 \eeq
 This potential is related by Fourier transformation to the amplitude of
$G-\widetilde G$ scattering via gluon exchange in the cross channel, as is
illustrated in Fig.~\ref{gg-scatt}.

Calculating the matrix elements with the operator 
Eq.~(\ref{2120})
between the states Eqs.~(\ref{2060}) - (\ref{2100}) we arrive at,
 \beqn
_S\la 8_a|\hat V(r)|8_{a'}\ra_S &=&
- \frac{21}{20}\,\frac{\alpha_s(r)}{r}\,
\delta_{aa'}\ ;
\label{2140}\\
_A\la 8_a|\hat V(r)|8_{a'}\ra_A &=&
- \frac{3}{2}\,\frac{\alpha_s(r)}{r}\,
\delta_{aa'}\ ;
\label{2160}\\
\la 1|\hat V(r)|1\ra &=& -3\,\frac{\alpha_s(r)}{r}\ ;
\label{2180}\\   
_S\la 8_a|\hat V(r)|8_{a'}\ra_A &=&
_S\la 8_a|\hat V(r)|1\ra =
_A\la 8_a|\hat V(r)|1\ra =0
\label{2200}   
\eeqn
 Here the subscripts $S$ and $A$ correspond to the symmetric and asymmetric states
Eqs.~(\ref{2060}) and (\ref{2080}) respectively.

Thus, the states Eqs.~(\ref{2060}) - (\ref{2100}) are the eigenstates of the
operator Eq.~(\ref{2120}). All the matrix elements have negative signs, what
means that also in color-octet states glue-glue interaction is attractive. We
assume that the relation between the nonperturbative potentials for the
glue-glue system in different states is the same as in the perturbative case. 
Then, we arrive to the following relation between the potentials
corresponding to the color octet and singlet states Eqs.~(\ref{2060}) -
(\ref{2100}),
 \beq 
\Re V_{8_S}\ :\ \Re V_{8_A}\ :\ \Re V_1\ =\ 
{7\over20}\ :\ {1\over2}\ :\ 1
\label{2220}   
 \eeq 
 Comparing with the known value of the parameter $b_0=0.65\GeV$ in the potential
Eq.~(\ref{2020}) for the color-singlet $|\bar ccG\ra$ state \cite{kst2} we get the
following values corresponding to fluctuations of a gluon containing a color-octet
$\bar cc$, symmetric and asymmetric, and one containing a color-singlet $\bar
cc$, Eq.~(\ref{2100}), 
 \beq
b_0 = \left\{ 
\begin{array}{ccc} 
0.55\GeV&\ \ \ \ \ {\rm for\ the\ symmetric\ octet-octet\ state\ Eq.}\ 
(\ref{2060})  \\
0.50\GeV&\ \ \ \ \ \ \ \ \,{\rm for\ the\ asymmetric\ octet-octet\ state\ Eq.}\ 
(\ref{2080})\\
0.20\GeV&\ \ \,{\rm for\ the\ singlet-octet\ state}\ .
\end{array}
\right.
\label{2240} 
 \eeq
 Thus, the strength of the interaction between the $\bar cc$ and gluon
varies from channel to channel. Correspondingly, the mean transverse
separation for the effective glue-glue dipole propagating through the nucleus
changes as well affecting the strength of gluon shadowing.

\subsection{Process-dependent gluon shadowing}\label{process-dep}

\begin{figure}[tbh]
\includegraphics{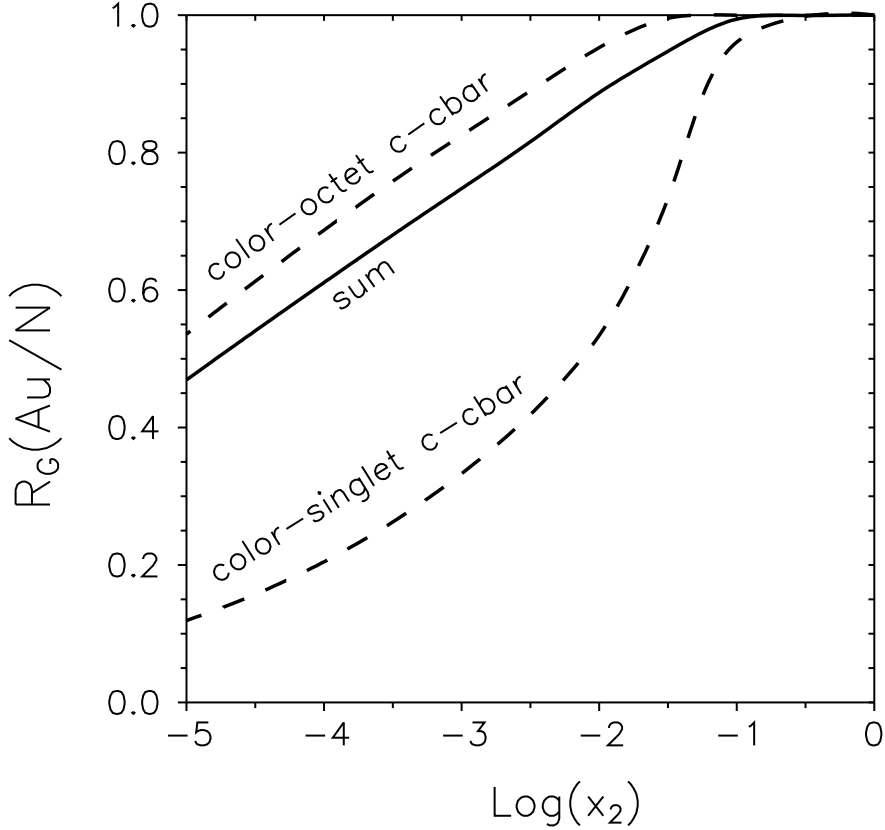}
\begin{center}
\vspace{11.5cm}
\parbox{13cm}
 {\caption[Delta]
 {\sl Ratio of gluon densities $R_G(Au/p)=G_{Au}(x_2)/195\,G_{p}(x_2)$
for color octet-octet and singlet-octet states $(\bar cc)-G$ (dashed curves).
The averaged gluon shadowing is depicted by solid curve.}
 \label{gl-shad}}
\end{center}
 \end{figure}

We calculated the gluon suppression factor in nuclei as function of $x_2$
according to Eq.~(\ref{800}) as,
 \beq
R_G(A/N) = 1-\frac{\Delta\sigma_G}
{A\,\sigma(GN\to \bar ccG\,X)}\ ,
\label{2260}
 \eeq using Eqs.~(\ref{820}) -- (\ref{1160}) and (\ref{2025}) for the
different values of $b_0$ Eq.~(\ref{2240}).  The results for gold are shown
by dashed curves in Fig.~\ref{gl-shad}. 
 The curves for the symmetric and antisymmetric states, Eqs.~(\ref{2060}) and
(\ref{2080}) are indistinguishable, since corresponding values of $b_0$ in
(\ref{2240}) are close. They are represented by a single (upper)  curve which
demonstrates a stronger shadowing effect than was found in \cite{kst2} for
gluon shadowing in DIS, since the value of $b_0$ is smaller. Gluon shadowing
corresponding to production of a colorless $\bar cc$ pair shown by the bottom
curve is the strongest one, as it was already found in \cite{kth} for
charmonium production.  Also the onset of shadowing happens differently, the
smaller $b_0$ is, the earlier (at larger $x_2$)  the onset of shadowing
occurs. This is because the mean effective mass of a $\bar ccG$ fluctuation
increases with $b_0$.

\subsection{\boldmath$D\bar D$ production from a color-singlet 
\boldmath$c\bar c$}\label{singlet}

Apparently, not every $\bar cc$ pair produced in a color-singlet state ends
up by formation of a charmonium. Some of them, in particular those which are
very heavy, can hadronize producing a $\bar DD$. Since the process of
color-singlet $\bar cc$ production is associated with a strongest effect of
gluon shadowing (see Fig.~\ref{gl-shad}) its contribution to open charm can
affect the expected nuclear shadowing. 

It is difficult to calculate reliably the fraction of colorless $\bar cc$
pairs decaying into open charm, since the hadronization dynamics is quite
complicated. One can rely on a simple recipe: the color-singlet pairs with
effective mass below or above the $\bar DD$ threshold end up with either
hidden or open charm respectively. 
Then, the fraction $\delta$ which goes to open charm is,
 \beq
\delta = \frac{
\int\limits_{4m_D^2}^\infty dM^2\,
{\frac{d\sigma}{dM^2}} }
{\int\limits_{4m_c^2}^{\infty}dM^2\,
\frac{d\sigma}{dM^2}}\ ,
\label{120}
 \eeq
 
The effective mass of a color-singlet pairs depends on relative transverse
momentum $\vec k_T$ and sharing of the longitudinal momentum $\alpha$,
 \beq
M_{\bar cc}^2 = \frac{m_c^2+k_T^2}{\alpha\,\bar\alpha}\ .
\label{b20} 
 \eeq
 Note that we must consider only the $\bar cc$ pairs which can be produced in
two-gluon fusion. Correspondingly, the mass distribution can be deduced from
the differential cross section which according to \cite{kst1,km} is given by
the double Fourier transform of the bilinear combination of the dipole
amplitudes,
 \beqn
&& \frac{d\sigma}{d^2k_T\,d\alpha} = 
\frac{\alpha_s}{32\,(2\pi)^4}
\int d^2r_1\,d^2r_2\,
e^{i\vec k_T\cdot(\vec r_1-\vec r_2)}
\nonumber\\ &\times&
\Bigl[m_c^2\,K_0(m_cr_1)\,K_0(m_cr_2) +
(\alpha^2+\bar\alpha^2)\,\vec\nabla_1 K_0(m_cr_1)\cdot
\vec\nabla_2 K_0(m_cr_2)\,\Bigr]
\nonumber\\ &\times&
\Bigl\{\sigma_{\bar qq}(\alpha\vec r_1 + \bar\alpha\vec r_2)
+\sigma_{\bar qq}(\alpha\vec r_2 + \bar\alpha\vec r_1)
- \sigma_{\bar qq}[\alpha(\vec r_1-\vec r_2)]
- \sigma_{\bar qq}[\bar\alpha(\vec r_1-\vec r_2)]
\Bigr\}
\label{b40}
 \eeqn
 Integration of this expression over $k_T$ and $\alpha$ recovers the cross
section Eq.~(\ref{410}) for production of color-singlet $G\,N\to (\bar
cc)_1\,X$.  The mean $\bar cc$ separation is small, hence one can employ the
dipole approximation, $\sig(r)=C\,r^2$.  Then the last factor in curly
brackets of (\ref{b40}) becomes as simple as $2\,C\,\vec r_1\cdot\vec r_2$,
and the cross section takes the form,
 \beq
\frac{d\sigma}{d^2k_T\,d\alpha} = 
\frac{\alpha_s\,C}{64\pi}\,
\left[\frac{4\,k_T^2\,m_c^2\,\alpha\,\bar\alpha}
{(k_T^2+m_c^2)^4} +
\frac{\alpha^2 + \bar\alpha^2}
{(k_T^2+m_c^2)^2}\right]\ ,
\label{60}
 \eeq
 which leads to the mass distribution,
 \beq
\frac{d\sigma}{dM^2} = 
\frac{\alpha_s\,C}{64\pi}\,
\int\limits_{1/2}^{\alpha_{max}}
d\alpha\,\left[\frac{4\,m_c^2}{M^6\,\alpha\,\bar\alpha} -
\frac{4\,m_c^4}{M^8\,\alpha^2\,\bar\alpha^2} +
\frac{\alpha^2+\bar\alpha^2}{\alpha\,\bar\alpha\,M^4}\right]\ ,
\label{80}
 \eeq
 where 
 \beqn
\alpha_{max} &=& {1\over2}\,\left(
1+\sqrt{1-v}\right)\nonumber\\
v &=&\frac{4m_c^2}{M^2}\ .
\label{100b}
 \eeqn

Eventually, the differential cross section can be represented as,
 \beq
\frac{d\sigma}{dM^2} = 
\frac{\alpha_s\,C}{64\pi\,M^4}\,
\left[\left(1+v-{1\over2}v^2\right)\,
\ln\left(\frac{1+\sqrt{1-v}}
{1-\sqrt{1-v}}\right)
 - (1+v)\sqrt{1-v}\right]\ .
\label{110b}
 \eeq
 With this mass distribution the fraction of the cross section above the
$\bar DD$ threshold calculated at $m_c=1.5\GeV$ turns out to be quite large,
 \beq
\delta = 0.86\ .
 \label{130b}
 \eeq
 We conclude that only a small fraction of colorless $\bar cc$ pairs
fragments into charmonia.

Apparently, the estimate Eq.~(\ref{130b}) is quite rough and gives only the
order of magnitude for the value in question. One can also try to evaluate
this number using available data for production rate for charmonia ($\Psi,\
\chi,\ \eta_c$) relative to open charm in high energy $pp$ collisions. We
accept that $J/\Psi$ yield is about $2.5\%$ of open charm \cite{dima}, and
that nearly $30\%$ of $J/\Psi$s originate from decays of $\chi_1$ and
$\chi_2$ \cite{chi}.  Further, according to the Landau-Yang theorem
production of $\chi_1$ is forbidden for collinear approximation in the parton
model. However, corrections due to nonzero longitudinal polarization of the
fusing gluons may be large \cite{andreas}. We rely on experimental value 0.7
\cite{chi,chi1-chi2} for the ratio of $\chi_1$ and $\chi_2$ production rates. 
Since the branchings for decay of $\chi_1$ and $\chi_2$ to $J/\Psi$ are
$27\%$ and $13.5\%$ respectively, we find that production rates of $\chi_{2}$
is approximately $1.5\%$ of open charm. The yield of the $S$-wave states,
$\eta_c$, can be estimated using the ratio $\eta_c$ to $\chi_2$ of production
rates which according to \cite{baier} depends on the wave functions of the
charmonia at the origin.  Using the wave functions from \cite{brodsky} we
found about three times more $\eta_c$s than $\chi_2$s.  Summing up the
channels which can originate from glue-glue fusion, $\chi$s and $\eta_c$, we
arrive at about $7\%$ of open charm, while $20\%$ of all $\bar cc$ pairs are
produced in color-singlet states. Then,
 \beq
\delta = 0.7\ .
\label{150b}
 \eeq

 Thus, both very different ways to evaluate $\delta$, show that the main
fraction of colorless $\bar cc$ pairs produced in $NN$ collisions hadronizes
to channels containing open charm. They also provide the scale of theoretical
uncertainty which is not too bad.  For further estimates we use the value
$\delta=0.75$.

We mix the octet and singlet contributions to gluon
shadowing with the weights found above, $R_G = (0.8\,R_G^8 +
0.14\,R_G^1)/0.94$. The result is shown in Fig.~\ref{gl-shad} by solid curve.
This effect is about twice as strong as gluon shadowing calculated in
\cite{kst2} for DIS demonstrating a substantial deviation from QCD
factorization. 

\section{Shadowing effects for charm production off nuclei}\label{charm-shad}

\begin{figure}[tbh]
\includegraphics{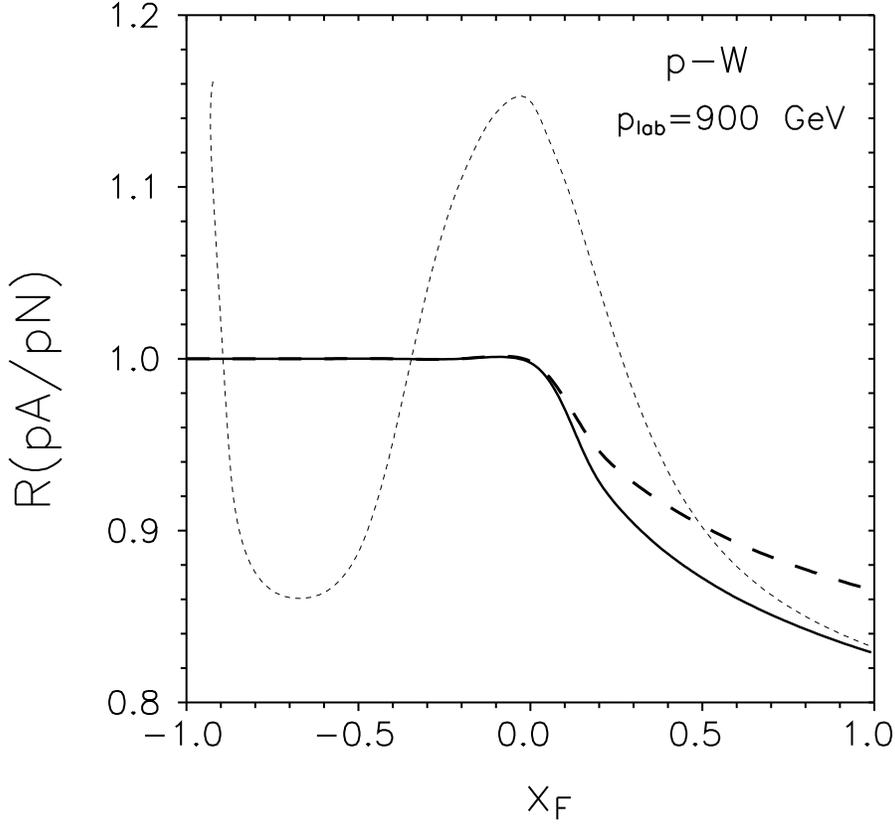}
\begin{center}
\vspace{11.5cm}
\parbox{13cm}
 {\caption[Delta]
 {\sl Nuclear effects for open charm production in $p_W$ collisions at
$900\GeV$ in the laboratory frame. The contribution of gluon shadowing is
shown by dashed curve. The solid curve represents the full shadowing effect
including shadowing of charmed quarks in the nucleus. Possible medium effects
including antishadowing and EMC-suppression for gluons \cite{eks} are also 
added and the result is represented by the dotted curve.}
 \label{pa-xf}}
\end{center}
 \end{figure}

To observe the shadowing effects in open charm production one must access the
kinematic region of sufficiently small $x_2 \lsim 0.1$. With fixed targets it
can be achieved at highest energies at Fermilab and in the experiment HERA-B
at DESY. We apply the results of previous section for gluon shadowing to
$\bar cc$ pair production in proton-nucleus collisions. We assume that the
$\bar cc$ is produced with Feynman $x_F$ corresponding to
$x_2=(-x_F+\sqrt{x_F^2+4M_{\bar cc}^2/s})/2$, where we fix $M_{\bar
cc}=4\GeV$. The contribution of gluon shadowing to nuclear effects in
proton-tungsten collision at $p_{lab}=900\GeV$ is depicted by the dashed
curve in Fig.~\ref{pa-xf}. 

The higher twist shadowing correction corresponding to the eikonalized dipole
cross section $\sigma_3$ in (\ref{500}) is also a sizeable effect and should
be added. It is diminished, however, due to the strong gluon shadowing which
also reduces the amount of gluons available for multiple interactions
compared to the eikonal approximation (\ref{500}). We take this reduction
into account multiplying $\sigma_3$ in (\ref{500})  by $R_G(x_2,M_{\bar
cc})$. The summed shadowing suppression of $\bar cc$ production is depicted
in Fig.~\ref{pa-xf} by the solid curve. 

Besides shadowing, other nuclear effect are possible. The EMC effect,
suppression of the nuclear structure function $F_2^A(x,Q^2)$ at large $x$, as
well as the enhancement at $x\sim 0.1$ should also lead to similar
modifications in the gluon distribution function $G^A(x,Q^2)$. These affects
are different from shadowing which is a result of coherence. A plausible
explanation relates them with medium effects, like swelling of bound nucleons
\cite{michele}. To demonstrate a possible size of the medium effects on gluon
distribution we parametrize and apply the effect of gluon enhancement and
suppression at large $x$ suggested in \cite{eks}. Although it is based on ad
hoc gluon shadowing and underestimated shadowing for valence quarks (see in
\cite{krtj}), it demonstrates the scale of possible effects missed in our
analysis. 

There are still other effects missed in our calculations.  At this energy the
effect of energy loss due to initial state interactions \cite{eloss} is
important enhancing nuclear suppression at large $x_F$ (compare with
\cite{kth}). Another correction is related to the observation that detection
of a charm hadron at large $|x_F|$ does not insure that it originates from a
charm quark produced perturbatively with the same $x_F$. Lacking gluons with
$x_{1,2}\to 1$ one can produce a fast charm hadron via a fast projectile
(usually valence) quark which picks up a charm quark created at smaller
$|x_F|$ (see also Sect.~\ref{low-energy}. This is actually the mechanism
responsible for the observed $D/\bar D$ asymmetry. It provides a rapidity
shift between the parent charm quark and the detected hadron.  Therefore, it
may reduce shadowing effects at largest $|x_F|$. We leave this problem open
for further study. 

\begin{figure}[tbh]
\includegraphics{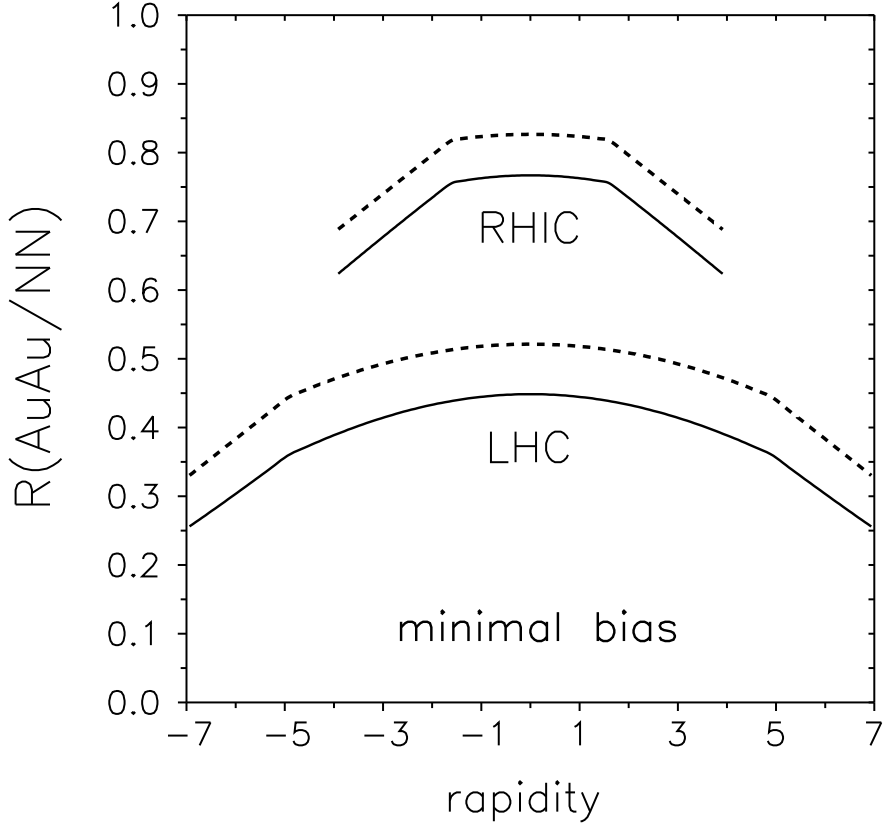}
\begin{center}
\vspace{11.5cm}
\parbox{13cm}
 {\caption[Delta]
 {\sl Nuclear shadowing for open charm production in minimal bias gold-gold
collision. Dotted curves correspond to net effect of gluon shadowing, while
solid curves include both effects of gluon shadowing and the higher twist
correction related to the nonzero separation of the $\bar cc$. The top (RHIC)
and bottom (LHC) curves correspond to $\sqrt{s}=200\GeV$ and $5500\GeV$
respectively.}
 \label{aa-mb}}
\end{center}
 \end{figure}

To predict shadowing effect in heavy ion collisions we employ QCD
factorization which we apply only for a given impact parameter. For minimal
bias events
 \beq
R_{AB}(y)=R_A(x_1)\,R_B(x_2)\ ,
\label{170b}
 \eeq
 where $y=\ln(x_1/x_2)/2$ is the rapidity of the $\bar cc$ pair.
Our predictions for RHIC ($\sqrt{s}=200\GeV$) and LHC ($\sqrt{s}=5500\GeV$)
are depicted in Fig.~\ref{aa-mb} separately for net gluon shadowing (dashed
curves) and full effect including quark shadowing (solid curves).
 Although shadowing of charmed quarks is a higher twist effect, its
contribution is about $10\%$ at RHIC and rises with energy.

One might be surprised by the substantial magnitude of shadowing expected at
the energy of RHIC. Indeed, the value of $x_{1,2}\approx 0.02$ at
mid-rapidity is rather large, and no gluon shadowing would be expected for
DIS \cite{kst2}.  However, the process of charm production demonstrates a
precocious onset of gluon shadowing as was discussed above.  Besides, the
nuclear suppression is squared in $AA$ collisions.

\begin{figure}[tbh]
\includegraphics{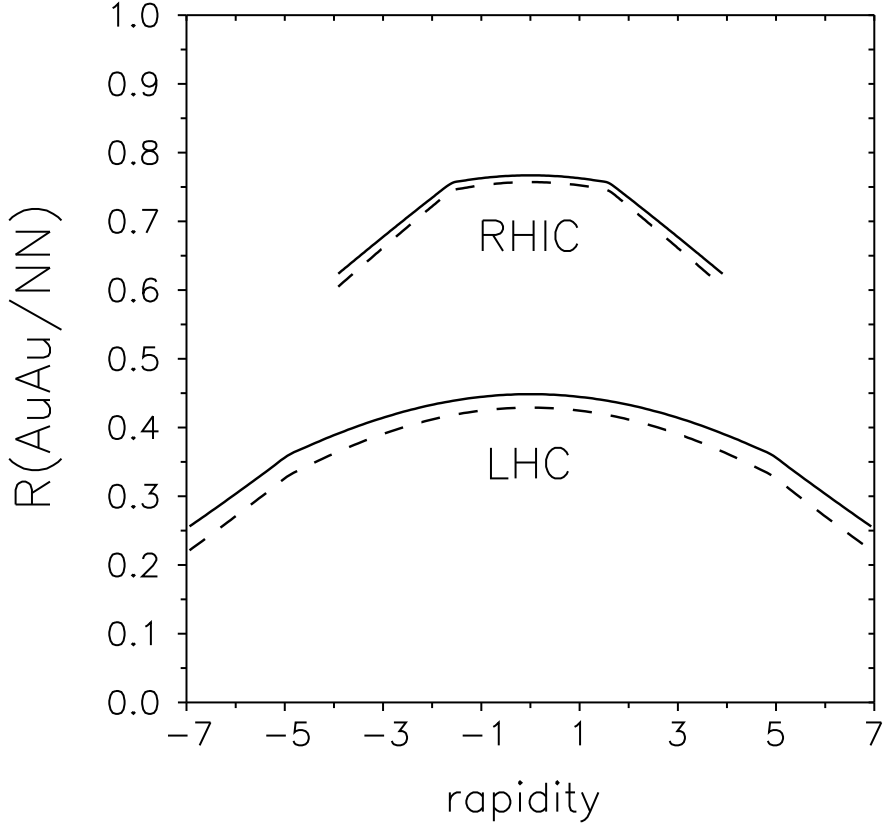}
\begin{center}
\vspace{11.5cm}
\parbox{13cm}
 {\caption[Delta]
 {\sl Comparison of minimal bias and central ($b=0$) collisions represented
by solid and dashed curves respectively. The top (RHIC) and bottom (LHC)
curves are calculated at $\sqrt{s}=200\GeV$ and $5500\GeV$ respectively. Both
effects of gluon shadowing and the higher twist correction are included.}
 \label{aa-centr}}
\end{center}
 \end{figure}

Unintegrated shadowing Eq.~(\ref{1060}) depends on impact parameter $\vec b$.
The strength of shadowing, $1-R_A(x,b)$, turns out to be nearly proportional
to nuclear thickness $T_A(b)$ \cite{knst,krtj}.  Nuclear shadowing for heavy
ion collision at impact parameter $b$ reads,
 \beq
R^{AB}(b)= \frac{1}{T_{AB}(b)}
\int d^2s\,R^A(s)\,T_A(s)\,R^B(\vec b-\vec s)\,
T_B(\vec b-\vec s)\ .
\label{bbb}
 \eeq
 This expression is normalized to one in absence of shadowing.
We calculated shadowing for most central collisions at $b=0$.
The results for the energies of RHIC and LHC are depicted by dashed curves in
Fig.~\ref{aa-centr} in comparison with shadowing for minimal bias collisions
shown by solid curves.
Calculations include both the gluon and quark shadowing effects.
 
Surprisingly, we hardly observe any difference between the expected shadowing
effects for minimal bias and central collisions. This contradicts a simple
intuition which relates central collisions to maximal nuclear thickness.  To
understand the source of such a similarity of the shadowing effects let us
consider a simple model for shadowing whose strength is proportional to the
full path length $L(b)$ in the nucleus (see the comment above), where
$L=2\sqrt{R_A^2-b^2}$ for the case of homogeneous nuclear density reads,
 \beq
1-R_A(x,b) = 
{2\over3}\,\Delta(x)\,\frac{L(b)}{R_A}\ .
\label{190b}
 \eeq
 Then, averaged over $b$ shadowing reads,
 \beq
R_A(x)= 1-\Delta(x)\ .
\label{200b}
 \eeq

 For minimal bias $AA$ collisions,
 \beq
R^{MB}_{AA}(y)= \Bigl[1 - \Delta(x_1)\Bigr]\,
\Bigl[1 - \Delta(x_2)\Bigr]\ .
\label{210b}
 \eeq

For central $AA$ collision at $b=0$, 
 \beq
R^{central}_{AA}(y) = 
\frac{1}{A\,\la T_A\ra}\int d^2s\ T_A^2(s)\,
\left[1 - {2\over3}\,\Delta(x_1)\,\frac{L(s)}{R_A}\right]\, 
\left[1 - {2\over3}\,\Delta(x_2)\,\frac{L(s)}{R_A}\right]\ .
\label{230b}
 \eeq
 With a constant nuclear density, $T_A(b)=2\rho_0\,L(b)$, we find,
 \beq
R^{central}_{AA}(y) =
1-{16\over15}\,[\Delta(x_1)+ \Delta(x_2)\Bigr] +
{32\over27}\,\Delta(x_1)\,\Delta(x_2)\ .
\label{240b}
 \eeq
 Comparing with shadowing for minimal bias events, Eq.~(\ref{210b}),
we see that the coefficients $16/15$ and $32/27$ in (\ref{240b})
are not very different from one. This is why $R^{MB}_{AA}(y)$
and $R^{central}_{AA}(y)$ depicted in Fig.~\ref{aa-centr} are so similar.

Thus, if data, like recent results from PHENIX \cite{phenix-charm},
demonstrate no difference in amount of produced charm between central and
minimal bias collisions of heavy ions, it should not be interpreted as an
indication to weak nuclear effects.  Indeed, Fig.~\ref{aa-centr} demonstrates
substantial nuclear shadowing effects at RHIC and especially at LHC, while
the difference between minimal bias and central collisions is tiny.  To
detect a signal of shadowing one should compare amount of produced charm in
central or minimal bias events with very peripheral, or better with $pp$
collisions. This is planned to be done at PHENIX \cite{phenix-charm}.

\section{Final state absorption of heavy flavored hadrons}
\label{low-energy}

One may expect no nuclear effects in the limit of short coherence length when
a heavy $\bar QQ$ pair is produced momentarily inside the nucleus and then
propagates through the nucleus.  No shadowing is possible in this case and
a question arises, if there is any nuclear suppression due to final state
interaction? On the contrary to a conventional expectation, the answer is
{\it yes}. 

Light hadrons are known to attenuate propagating through a nuclear
matter. Although high energy quarks cannot be absorbed or stopped via
soft interactions, any inelastic collision, like $M\,N\to M\,X$ leads to
softening of the momentum spectrum of the meson. This is known from data
and can be interpreted as a color-exchange collision which leads to
independent hadronization of the projectile valence $q$ and $\bar q$
which carry parts of the total initial meson momentum. The newly produced
meson can carry only a fraction of the $q$ or $\bar q$ momenta, in
average about a half of the initial meson momentum. Therefore, if one
detects a meson with large $x_F$ any inelastic interaction of the
produced meson on its way out of the nucleus shifts its momentum down to
smaller $x_F$ what looks like absorption. The same obviously is valid for
light baryons. Thus, nuclear attenuation of light hadrons can be and has
been widely used as a source of information about the hadron-nucleon
inelastic cross sections. Such information is especially precious and
unique for unstable hadrons.

At first glance, one may expect a different situation for heavy flavored
hadrons. If a hadron contains only one heavy quark (open flavor), even
simple kinematic consideration shows that the heavy quark carries the
major fraction of the hadron momentum. Therefore, after the hadron
breaks up in an inelastic collision, the heavy quark picks up an
antiquark from vacuum and the recovered heavy flavored baryon has nearly
the same momentum as its predecessor. Thus, one might think that hadrons
with open heavy flavor propagate through nuclear matter practically without
attenuation even at large $x_F$ (except the very endpoint values $x_F >
1-m_q/m_Q$).

However, a deeper insight shows that such a consideration is oversimplified.
Indeed, a heavy quark pair $\bar QQ$ is produced mainly in central rapidity
region at $x_F\ll 1$. In order to become a constituent of an energetic heavy
hadron, the $Q$ or $\bar Q$ has to be picked up by the energetic valence
quarks of the beam hadron. Therefore, the typical configuration in which the
hadron with open heavy flavor is produced, is very different from one
discussed above. This is not the hadron yet, but just a colorless light-heavy
quark system which needs time (formation length) to develop the wave function
of the hadron,
 \beq
l_f \sim \frac{E_H}{2\,\omega\,m_H}\ ,
\label{1000b}
 \eeq
 where $E_H,\ m_H$ are the hadron energy and mass, and $\omega\sim 300\,\MeV$
is the oscillator frequency. Thus, at high energies, $E_H > 10\,\GeV$ (for
charm)  this produced colorless quark system propagates through the nucleus
in such an unusual configuration and forms the hadronic wave function far
away from the nucleus.  At the same time, the energy should not be too high
to guarantee that the coherence time $t_c=2E_{\bar QQ}/M^2_{\bar QQ} \ll
R_A$. 

What happens if such a quark configuration, in which the heavy quark
carries a small fraction of the total momentum, interacts inelastically
in the nuclear medium? The answer is opposite to our previous
expectation. The new heavy flavored hadron will be produced via
hadronization of the slow heavy quark, therefore will be removed by the
inelastic interaction from the high $x_F$ region. We conclude that
produced heavy flavored hadrons are effectively absorbed the same way as
light hadrons.  This absorption leads to a suppression factor,
 \beq
S_A^{FSI}=\frac{1}{\sigma^{HN}_{in}}
\int d^2b\left[1 - 
e^{-\sigma^{HN}_{in}T_A(b})\right]\ ,
\label{1010b}
 \eeq
 where $\sigma^{HN}_{in}$ is the cross section of inelastic interaction of
the pre-formed heavy flavored hadron $H$ with a nucleon.  This suppression
factor can be applied only at large $x_F$. At small $x_F\ll 1$ this
production mechanism is replaced by an independent fragmentation of the
created heavy quarks into the hadron $H$. Such a hadron should be produced in
the conventional configuration when the heavy quark carries the main fraction
of the momentum. Therefore, we expect no nuclear suppression at small $x_F$
as far as shadowing is absent. There might be even an enhancement due to
feeding from the large $x_F$ events, however this effect should be very small
because the $x_F$ distribution of $H$ steeply falls down towards $x_F=1$. 

Note that measurement of the suppression factor (\ref{1010b}) provides an
access to precious information about the interaction cross section of heavy
flavored hadrons. The best energy range for such study would be a few tens of
GeV (for charm). At the low end of this energy interval, $E_H \sim 10\GeV$
one can additionally study the variation of the formation length with the
energy of the hadron. No absorption is expected at $l_f\to 0$, while
absorption increase with energy and saturates at $l_f\gg R_A$. 

A very interesting study of interaction of the formed H states with a
nucleons can be done using an antiproton beam with a high momentum
resolution. In the reaction $\bar p\,p \to \bar H\,H$ both hadrons are
produced with the maximal momentum $x_F=1$. At $E_{\bar p} \lsim 10\GeV$
the wave function of $\bar H$ is formed shortly and then any inelastic
interaction of $\bar H$ leads to the suppression factor
Eq.~(\ref{1010b})\footnote{In addition, energy dependence of the reactions
$\bar p\,p \to D\,\bar D$ and $\bar p\,p \to \bar B_c\, B_c$ would
provide an important information about the intercepts of Regge
trajectories associated with $B_c$ and $D$ states, respectively.}.
Thus, in this reaction one is uniquely able to measure the genuine $HN$ 
($H=D,\,B_c$) inelastic cross section.

\section{Conclusions}\label{summary}

Production of heavy flavored hadrons off nuclei is subject to nuclear
shadowing. QCD factorization predicts it to be related to gluon shadowing,
while the higher twist correction suppressed by the heavy quark mass squared
should be neglected. However, gluon shadowing is still unknown from data and
nuclear effects for heavy flavor production could not be predicted so far. 

In this paper we extend the light-cone dipole phenomenology to the case of
open charm production. This approach has been previously developed and
successfully applied to charmonium production \cite{kth}, Drell-Yan process
\cite{hir,krtj,gay,joerg}, photoproduction of vector mesons \cite{hikt,knst}
and deep-inelastic scattering \cite{gbw}. The advantage of this phenomenology
is possibility to use the universal dipole cross section fitted to the vast
amount of DIS data \cite{gbw}. This approach allows not only predict gluon
shadowing including its dependence on impact parameter, but also calculate
the higher twist correction related to shadowing of heavy quarks.  In every
case when predictions of this approach can be checked with available data, a
good agreement is achieved. 

Strong nonperturbative effects modifying the light-cone wave function of Fock
states containing gluons were found in \cite{kst2}. These effects
substantially suppress the cross section of diffractive gluon radiation and
gluon shadowing in nuclei. The interaction between partons propagating along
the light-cone naturally depends on the color state of the partons. Thus, we
predict a process dependence for gluon shadowing which is a manifestation of
deviation from QCD factorization. In particular, we expect about twice as
strong gluon shadowing for $\bar cc$ pair production off nuclei as for DIS,
which is, however, weaker than for production of charmonia.  The standard
parton model has no tool to access this strong deviation from QCD
factorization. Predictions \cite{ekv} for open charm production off nuclei
are based on the ad hoc gluon shadowing of \cite{eks}. However, even if DIS
data on nuclei were sufficiently precise to find gluon shadowing, one should
not apply the result to open charm production which is subject to much
stronger shadowing.

We predict substantial shadowing effects for gluons and charmed quarks in
heavy nuclei to be tested by future measurements at HERA-B, RHIC and LHC. In
particular, we studied centrality dependence of these effects and found
almost no difference between central and minimal bias collisions of heavy
ions. Such a similarity indeed was observed in the PHENIX experiment at RHIC
recently. However, it should not be interpreted as weakness of the shadowing
effect, on the contrary, we expect shadowing to be rather strong. 

We also considered the limit of short coherence length which takes place at
medium energies. Although no shadowing is possible in this regime, we expect
strong absorption effects due to final state interaction. This is a
consequence of an unusual configuration in which the heavy flavored hadron is
originally produced.

\bigskip 
 \noindent {\bf Acknowledgment}: We are grateful to J\"org H\"ufner 
read the manuscript and made many valuable comments, and Andreas Sch\"afer
for numerous inspiring and informative discussions. This work has been
supported by the grant from the Gesellschaft f\"ur Schwerionenforschung
Darmstadt (GSI), grant No.~GSI-OR-SCH, and by the grant INTAS-97-OPEN-31696.
An essential part of this work has been performed while A.V.T. was employed
by Regensburg and Heidelberg Universities. The works was finished while
B.Z.K. was visiting at ITP, Santa Barbara, whose support under NSF Grant
PHY99-07949 and hospitality are kindly acknowledged.

 
 \def\appendix{\par
 \setcounter{section}{0}
\setcounter{subsection}{0}
 \def\thesection{Appendix \Alph{section}}
\def\thesubsection{\Alph{section}.\arabic{subsection}}
\def\theequation{\Alph{section}.\arabic{equation}}
\setcounter{equation}{0}}

 \appendix
 
\section{Dipole representation for $\bar ccG$ production}
\label{diagrams}
\setcounter{equation}{0}

\begin{figure}[tbh]
\includegraphics{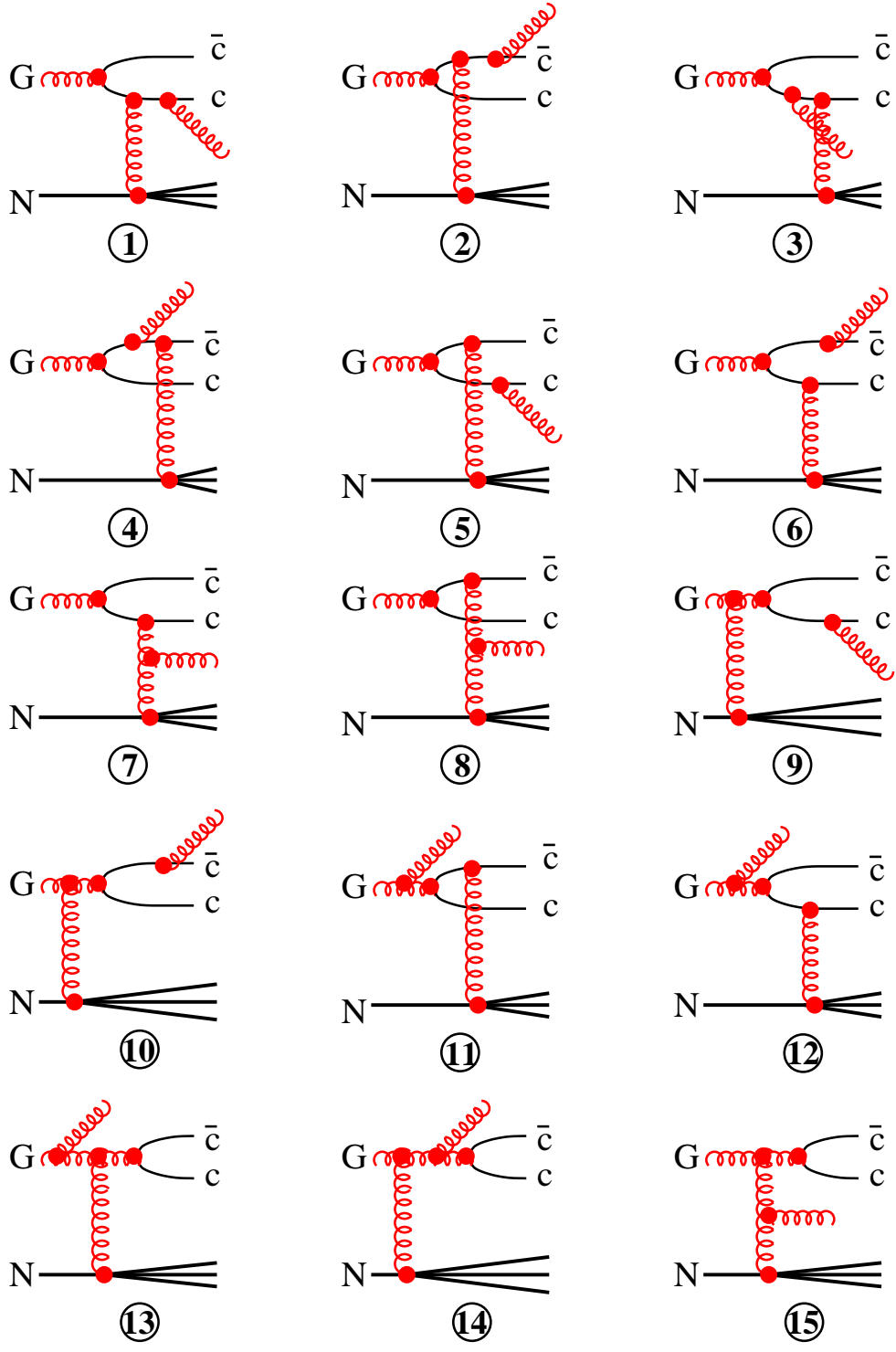}
\begin{center}
\vspace{19cm}
\parbox{13cm}
{\caption[Delta]
{\sl Born graphs contributing to $\bar cc$ pair production accompanied with 
radiation of a gluon.}
\label{graphs}}
\end{center}
 \end{figure}

The amplitude of the process
 \beq
G_a + N \to \bar cc + G_b + X\ ,
\label{d0}
 \eeq
 where $G_{a,b}$ are gluons
in color states $a$ and $b$, is described in Born
approximation by the set of 15 Feynman graphs depicted in
Fig.~\ref{graphs}.  

The amplitude of $\bar ccG$ production corresponding to 
graphs 1-15 has the following structure,
 \beq
A^{\bar \mu\mu}_{ij,ab} = 
\sum\limits_{l=1}^{15} 
\Bigl(A^{\bar \mu\mu}_{ij,ab}\Bigr)_l\ ,
\label{d1}
 \eeq
 where
 \beq
\Bigl(A^{\bar \mu\mu}_{ij,ab}\Bigr)_l = 
\frac{i\,\sqrt{3}\,\alpha_s^{3/2}}
{(k_T^2+\lambda^2)\,D_l}\ 
\sum\limits_{d=1}^{N_c^2-1} 
T_{abd}^{(l)}(ij)\,
F^{(d)}_{GN\to X}(k_T,\{X\})\ 
\xi^\mu_c\Bigr.^\dagger\,\hat\Gamma_l\,
\bar\xi^{\bar\mu}_{\bar c}\ .
\label{d2}
 \eeq
 Here $\lambda$ is the effective gluon mass aimed to incorporate
confinement, its value we discuss later.  The amplitude of gluon-absorption
by a nucleon, $F^{(d)}_{GN\to X}(k_T,\{X\})$, determines the unintegrated
gluon density as it is introduced in (\ref{390}),
 \beq
\int d\{X\}\sum\limits_{d=1}^8 
\Bigl|F^{(d)}_{GN\to X}(k_T,\{X\})\Bigr|^2 =
4\pi\,{\cal F}(k_T^2,x)\ .
\label{d2a}
 \eeq
 Here
 \beqn
x &=& \frac{M^2(\bar c,c,G)}{s}\ ;
\label{d2b}\\
M^2_{\bar ccG} &=&
\frac{m_c^2+k_1^2}{\alpha_1} +
\frac{m_c^2+k_2^2}{\alpha_2} +
\frac{\lambda^2+k_3^2}{\alpha_3}\ ,
\label{d2c}
 \eeqn
where $\vec k_1,\ \vec k_2,\ \vec k_3$ and $\alpha_1,\
\alpha_2,\ \alpha_3$ are the transverse momenta and fractions of
the initial light-cone momentum of the projectile gluon carried by
the produced $\bar c,\ c$ and $G$ (see Fig.~\ref{1graph}),
respectively, and
 \beq
\vec k_T =
\vec k_1 +\vec k_2 +\vec k_3\ .
\label{d2d} 
 \eeq

The $3\times3$ matrixes $T_{abd}^{(l)}(ij)$ in (\ref{d2}) act in the
color space of the $\bar cc$ quarks, and the indexes $i,\ j$
corresponds to the $c$ and $\bar c$ respectively.
 \beqn
T^{(1)}_{abd} &=&
\tau_a\,\tau_d\,\tau_b\nonumber\\
T^{(2)}_{abd} &=&
\tau_b\,\tau_d\,\tau_a\nonumber\\
T^{(3)}_{abd} &=&
\tau_a\,\tau_b\,\tau_d\nonumber\\
T^{(4)}_{abd} &=&
\tau_d\,\tau_b\,\tau_a\nonumber\\
T^{(5)}_{abd} &=&
\tau_d\,\tau_a\,\tau_b\nonumber\\
T^{(6)}_{abd} &=&
\tau_b\,\tau_a\,\tau_d\nonumber\\
T^{(7)}_{abd} &=& 
i\,\sum\limits_{e=1}^{N_c^2-1}f_{ebd}\,
\tau_a\,\tau_e\ ;\nonumber\\
T^{(8)}_{abd} &=& 
i\,\sum\limits_{e=1}^{N_c^2-1}f_{edb}\,
\tau_e\,\tau_a\ ;\nonumber\\
T^{(9)}_{abd} &=& 
i\,\sum\limits_{e=1}^{N_c^2-1}f_{ade}\,
\tau_e\,\tau_b\ ;\nonumber\\
T^{(10)}_{abd} &=& 
i\,\sum\limits_{e=1}^{N_c^2-1}f_{ade}\,
\tau_b\,\tau_e\ ;\nonumber\\
T^{(11)}_{abd} &=& 
i\,\sum\limits_{e=1}^{N_c^2-1}f_{abe}\,
\tau_e\,\tau_d\ ;\nonumber\\
T^{(12)}_{abd} &=& 
i\,\sum\limits_{e=1}^{N_c^2-1}f_{abe}\,
\tau_d\,\tau_e\ ;\nonumber\\
T^{(13)}_{abd} &=& 
i\,\sum\limits_{e,g=1}^{N_c^2-1}f_{abe}\,
f_{edg}\,\tau_g\ ;\nonumber\\
T^{(14)}_{abd} &=& 
i\,\sum\limits_{e,g=1}^{N_c^2-1}f_{ade}\,
f_{ebg}\,\tau_g\ ;\nonumber\\
T^{(15)}_{abd} &=& 
i\,\sum\limits_{e,g=1}^{N_c^2-1}f_{ebd}\,
f_{aeg}\,\tau_g\ .
\label{a10}
 \eeqn
 Here $\lambda_a = \tau_a/2$ are the Gell-Mann matrixes.

Note that the matrixes $T^{(l)}_{abd}$ are not independent, but 
connected by linear relations (we skip the indexes $abd$),
 \beqn 
T^{(3)} - T^{(1)} + T^{(7)} &=& 0\ ;\nonumber\\ 
T^{(5)} - T^{(1)} + T^{(9)} &=& 0\ ;\nonumber\\ 
T^{(4)} - T^{(2)} + T^{(8)} &=& 0\ ;\nonumber\\ 
T^{(6)} - T^{(2)} + T^{(10)} &=& 0\ ;\nonumber\\ 
T^{(13)} - T^{(11)} + T^{(13)} &=& 0\ ;\nonumber\\ 
T^{(13)} + T^{(14)} + T^{(15)} &=& 0\ ;\nonumber\\ 
T^{(15)} - T^{(7)} + T^{(8)} &=& 0\ ;\nonumber\\ 
T^{(14)} - T^{(10)} + T^{(9)} &=& 0\ ;\nonumber\\ 
\label{a20}
 \eeqn                                                                    

The $c$-quark spinors $\xi$ in (\ref{d2}) are defined in
(\ref{370}); $\{X\}$ is the set of variables describing the state
$X$;  the 15 vertex functions $\hat\Gamma_l$ read,
 \beqn
\hat\Gamma_1 &=&
\hat U_1(\vec k_1,\alpha_1)\, 
\hat V_1(\vec k_{23},\alpha_2,\alpha_3)
\ ;\nonumber\\
\hat\Gamma_2 &=& 
\hat V_2(\vec k_{13},\alpha_1,\alpha_3)\,
\hat U_2(\vec k_2,\alpha_2)\, 
\ ;\nonumber\\
\hat\Gamma_3 &=& - 
\alpha_1\,\hat U_1(\vec k_1,\alpha_1)\, 
\hat V_1(\vec k_{23}-\alpha_3\vec k_T,\alpha_2,\alpha_3)
\ ;\nonumber\\
\hat\Gamma_4 &=& - \alpha_2\,
\hat V_2(\vec k_{13}-\alpha_3\vec k_T,\alpha_1,\alpha_3)\,
\hat U_2(\vec k_2,\alpha_2)\, 
\ ;\nonumber\\
\hat\Gamma_5 &=& - \alpha_2\alpha_3\,
\hat U_1(\vec k_1 - \vec k_T,\alpha_1)\, 
\hat V_1(\vec k_{23},\alpha_2,\alpha_3)
\ ;\nonumber\\
\hat\Gamma_6 &=& -\alpha_1\alpha_3\, 
\hat V_2(\vec k_{13},\alpha_1,\alpha_3)\,
\hat U_2(\vec k_2 - \vec k_T,\alpha_2)\, 
\ ;\nonumber\\
\hat\Gamma_7 &=& - \alpha_1\,
\hat U_1(\vec k_1,\alpha_1)\, 
\hat V_1(\vec k_{23} + \alpha_2\vec k_T,\alpha_2,\alpha_3)
\ ;\nonumber\\
\hat\Gamma_8 &=& - \alpha_2\,
\hat V_2(\vec k_{13} + \alpha_1\vec k_T,\alpha_1,\alpha_3)\,
\hat U_2(\vec k_2,\alpha_2)\, 
\ ;\nonumber\\
\hat\Gamma_9 &=& - \alpha_2\alpha_3\,
\hat U_1(\vec k_1 - \alpha_1\vec k_T,\alpha_1)\, 
\hat V_1(\vec k_{23},\alpha_2,\alpha_3)
\ ;\nonumber\\
\hat\Gamma_{10} &=& - \alpha_1\alpha_3\, 
\hat V_2(\vec k_{13},\alpha_1,\alpha_3)\,
\hat U_2(\vec k_2 - \alpha_2\vec k_T,\alpha_2)
\ ;\nonumber\\
\hat\Gamma_{11} &=& \alpha_3\,
\hat U_0(\vec k_{12} + \alpha_1\vec k_T,\alpha_1,\alpha_2)\, 
\hat V_0(\vec k_{3})
\ ;\nonumber\\
\hat\Gamma_{12} &=& - \alpha_3\,
\hat U_0(\vec k_{12} - \alpha_2\vec k_T,\alpha_1,\alpha_2)\, 
\hat V_0(\vec k_3)
\ ;\nonumber\\
\hat\Gamma_{13} &=& \bar\alpha_3\,
\hat U_0(\vec k_{12},\alpha_1,\alpha_2)\, 
\hat V_0(\vec k_3)
\ ;\nonumber\\
\hat\Gamma_{14} &=& \alpha_1\alpha_2\,
\hat U_0(\vec k_{12},\alpha_1,\alpha_2)\, 
\hat V_0(\vec k_3)\ ;\nonumber\\
\hat\Gamma_{15} &=& \alpha_1\alpha_2\,
\hat U_0(\vec k_{12},\alpha_1,\alpha_2)\, 
\hat V_0(\vec k_3-\vec k_T)\ .
\label{d3}
 \eeqn 
 Here 
 \beqn 
&&\vec k_{13} = 
\alpha_3\vec k_1 - \alpha_1\vec k_3\ ; 
\nonumber\\ 
&&\vec k_{23} =
\alpha_3\vec k_2 - \alpha_2\vec k_3\ ; 
\nonumber\\ 
&&\alpha_1+\alpha_2+\alpha_3 = 1\ . 
\label{d3a}
 \eeqn                                                           

The matrixes $\hat U_{0,1,2}$ and $\hat V_{0,1,2}$ are defined as
follows,
 \beqn
\hat U_0(\vec k_{12},\alpha_1,\alpha_2) &=& 
(\alpha_1+\alpha_2)m_c\, \vec\sigma\cdot\vec e +
(\alpha_2-\alpha_1)\,(\vec\sigma\cdot\vec n)
(\vec e_{in}\cdot\vec k_{12}) + 
i\,(\vec e_{in}\times\vec n)\cdot\vec k_{12}\ ;
\nonumber\\
\hat U_1(\vec k_1,\alpha_1) &=& 
m_c\, \vec\sigma\cdot\vec e +
(1-2\alpha_1)\,(\vec\sigma\cdot\vec n)
(\vec e_{in}\cdot\vec k_1) + 
i\,(\vec e_{in}\times\vec n)\cdot\vec k_1\ ;
\nonumber\\
\hat U_2(\vec k_2,\alpha_2) &=& 
m_c\, \vec\sigma\cdot\vec e +
(1-2\alpha_2)\,(\vec\sigma\cdot\vec n)
(\vec e_{in}\cdot\vec k_2) - 
i\,(\vec e_{in}\times\vec n)\cdot\vec k_2\ ;
\label{d4a}\\
\hat V_0(\vec k_3) &=& \vec e_f\cdot\vec k_3
\ ;\nonumber\\ 
\hat V_1(\vec k_{23},\alpha_2,\alpha_3) &=&
(2\alpha_2+\alpha_3)\,\vec k_{23}\cdot\vec e_f +
i\,m_c\,\alpha_3^2\,(\vec n\times\vec e_f)\cdot\vec\sigma -
i\,\alpha_3\,(\vec k_{23}\times\vec e_f)\cdot\vec\sigma\ ;
\nonumber\\
\hat V_2(\vec k_{13},\alpha_1,\alpha_3) &=&
(2\alpha_1+\alpha_3)\,\vec k_{13}\cdot\vec e_f -
i\,m_c\,\alpha_3^2\,(\vec n\times\vec e_f)\cdot\vec\sigma +
i\,\alpha_3\,(\vec k_{13}\times\vec e_f)\cdot\vec\sigma\ ,
\nonumber\\
\label{d4b}
\eeqn
 where $\vec e_{in}$ and $\vec e_f$ are the polarization vectors of the 
initial and radiated gluons respectively.

The functions $D_l$ in the denominator of (\ref{d2}) read,
 \beqn
D_1 &=& \Delta_0(\vec k_1,\alpha_1)\,
\Delta_2(\vec k_{23},\alpha_2,\alpha_3)\ ;
\nonumber\\
D_2 &=& \Delta_0(\vec k_2,\alpha_2)\,
\Delta_2(\vec k_{13},\alpha_1,\alpha_3)\ ;
\nonumber\\
D_3 &=& \Delta_0(\vec k_1,\alpha_1)\,
\Delta_1(\vec k_1,\vec k_{23} - 
\alpha_3\vec k_T,\alpha_1,\alpha_2,\alpha_3)\ ;
\nonumber\\
D_4 &=& \Delta_0(\vec k_2,\alpha_2)\,
\Delta_1(\vec k_2,\vec k_{13} - 
\alpha_3\vec k_T,\alpha_2,\alpha_1,\alpha_3)\ ;
\nonumber\\
D_5 &=& \Delta_1(\vec k_1 - \vec k_T,
\vec k_{23},\alpha_1,\alpha_2,\alpha_3)\,
\Delta_2(\vec k_{23},\alpha_2,\alpha_3)\ ;
\nonumber\\
D_6 &=& \Delta_1(\vec k_2 - \vec k_T,
\vec k_{13},\alpha_2,\alpha_1,\alpha_3)\,
\Delta_2(\vec k_{13},\alpha_1,\alpha_3)\ ;
\nonumber\\
D_7 &=& \Delta_0(\vec k_1,\alpha_1)\,
\Delta_1(\vec k_1,\vec k_{23} + \alpha_2\vec k_T,
\alpha_1,\alpha_2,\alpha_3)\ ;
\nonumber\\
D_8 &=& \Delta_0(\vec k_2,\alpha_2)\,
\Delta_1(\vec k_2,\vec k_{13} + \alpha_1\vec k_T,
\alpha_2,\alpha_1,\alpha_3)\ ;
\nonumber\\
D_9 &=& \Delta_2(\vec k_{23},\alpha_2,\alpha_3)\,
\Delta_1(\vec k_1 - \alpha_1\vec k_T,\vec k_{23},
\alpha_1,\alpha_2,\alpha_3)\ ;
\nonumber\\
D_{10} &=& \Delta_2(\vec k_{13},\alpha_1,\alpha_3)\,
\Delta_1(\vec k_2 - \alpha_2\vec k_T,\vec k_{13},
\alpha_2,\alpha_1,\alpha_3)\ ;\nonumber\\ 
D_{11} &=& \Delta_3(\vec k_{3},\alpha_1,\alpha_3)\,
\Delta_4(\vec k_3,\vec k_{12} + 
\alpha_1\vec k_T,\alpha_1,\alpha_2,\alpha_3)
\ ;\nonumber\\ 
D_{12} &=& \Delta_3(\vec k_{3},\alpha_1,\alpha_3)\,
\Delta_4(\vec k_3,\vec k_{12} - \alpha_2\vec 
k_T,\alpha_1,\alpha_2,\alpha_3)
\ ;\nonumber\\ 
D_{13} &=& \Delta_3(\vec k_{3},\alpha_1,\alpha_3)\,
\Delta_5(\vec k_{12},\alpha_1,\alpha_2,\alpha_3)
\ ;\nonumber\\ 
D_{14} &=& \Delta_5(\vec k_{12},\alpha_1,\alpha_2,\alpha_3)\,
\Delta_4(\vec k_{31} + \vec k_{32},\vec 
k_{12},\alpha_1,\alpha_2,\alpha_3)
\ ;\nonumber\\ 
D_{15} &=& \Delta_5(\vec k_{12},\alpha_1,\alpha_2,\alpha_3)\,
\Delta_4(\vec k_3 - \vec k_T,\vec 
k_{12},\alpha_1,\alpha_2,\alpha_3)\ ,
\label{d5}
 \eeqn
 where
 \beqn
\Delta_0(\vec k_1,\alpha_1) &=&
m_c^2 + k_1^2 - \alpha_1\bar\alpha_1\,\lambda^2\ ;
\nonumber\\
\Delta_1(\vec k_1,\vec k_{23},\alpha_1,\alpha_2,\alpha_3) &=& 
\bar\alpha_1\bar\alpha_3(\alpha_3\,m_c^2 + 
\alpha_1\alpha_2\,\lambda^2) +
\alpha_1k_{23}^2 + \alpha_2\alpha_2\,k_1^2\ ;
\nonumber\\
\Delta_2(\vec k_{13},\alpha_1,\alpha_3) &=&
\alpha_3^2\,m_c^2 + 
\alpha_1(\alpha_1+\alpha_3)\,\lambda^2 + 
k_{13}^2\ ;\nonumber\\
\Delta_3(\vec k_3,\alpha_1,\alpha_3) &=&
k_3^2 + [1-\alpha_3(\alpha_1+\alpha_1)]\,\lambda^2\ ; 
\nonumber\\
\Delta_4(\vec k_3,\vec k_{12},\alpha_1,\alpha_2,\alpha_3) 
&=& \bar\alpha_3(\alpha_3\,m_c^2 + 
\alpha_1\alpha_2\,\lambda^2) +
\frac{\alpha_1\alpha_2}{\bar\alpha_3}\,k_{3}^2 + 
\frac{\alpha_3}{\bar\alpha_3}\,k_{12}^2\ ;
\nonumber\\
\Delta_5(\vec k_{12},\alpha_1,\alpha_2,\alpha_3) &=&
\bar\alpha_3^2\,m_c^2 - 
\alpha_1\alpha_2\,\lambda^2 + 
k_{12}^2\ .
\label{d6}
 \eeqn

Functions $\Delta_0$ - $\Delta_5$ are not independent, but 
satisfy the relation
 \beqn
&&\Bigl[\Delta_0(\vec k_1,\alpha_1)\,
\Delta_2(\vec k_{23},\alpha_2,\alpha_3)\Bigr]^{-1} =
\alpha_1\,\Bigl[\Delta_0(\vec k_1,\alpha_1)\,
\Delta_1(\vec k_1,\vec k_{23},\alpha_1,\alpha_2,\alpha_3)
\Bigr]^{-1}
\nonumber\\ &+&
\alpha_2\alpha_3\,
\Bigl[\Delta_2(\vec k_{23},\alpha_2,\alpha_3)\,
\Delta_1(\vec k_1,\vec k_{23},\alpha_1,\alpha_2,\alpha_3)
\Bigr]^{-1}\ ;\label{d7a}\\
&&\Bigl[\Delta_3(\vec k_3,\alpha_1,\alpha_3)\,
\Delta_5(\vec k_{12},\alpha_1,\alpha_2,\alpha_3)\Bigr]^{-1} 
=
\frac{\alpha_1\alpha_2}{\bar\alpha_3}\nonumber\\
&\times&\Bigl[\Delta_4(\vec k_3,\vec 
k_{12},\alpha_1,\alpha_2,\alpha_3)\,
\Delta_5(\vec k_{12},\alpha_1,\alpha_2,\alpha_3)
\Bigr]^{-1}
\nonumber\\ &+&
\frac{\alpha_3}{\bar\alpha_3}\,
\Bigl[\Delta_4(\vec k_3,\vec k_{12},\alpha_1,\alpha_2,\alpha_3)\,
\Delta_3(\vec k_3,\alpha_1,\alpha_3)
\Bigr]^{-1}\ .
\label{d7b}
 \eeqn  

The sum of the 15 amplitudes Eq.~(\ref{d2}) is convenient to split
up into 9 terms,
 \beq
\hat M = \sum\limits_{l=1}^{15}
T^{(l)}\,\frac{\hat\Gamma_l}{D_l} 
= \sum\limits_{i=1}^{9}S^{(i)}\,\hat M_i\ ,
\label{d8} 
 \eeq
 where matrixes $S^{(i)}$ are related to $T^{(l)}$ in (\ref{a10}) as,
 \beqn 
S^{(1)} &=& T^{(1)}\ ;\ \ \ \ \ \ \ S^{(2)} =T^{(2)}\
;\nonumber\\ 
S^{(3)} &=& T^{(7)}\ ;\ \ \ \ \ \ \ S^{(4)} =T^{(8)}\
;\nonumber\\ 
S^{(5)} &=& T^{(9)}\ ;\ \ \ \ \ \ \ S^{(6)} =T^{(10)}\
;\nonumber\\ 
S^{(7)} &=& {1\over2}\, \Bigl(T^{(11)} + T^{(12)}\Bigr)\
;\nonumber\\ 
S^{(8)} &=& T^{(14)}\ ;\ \ \ \ \ \ \ S^{(9)} =T^{(15)}\ . 
\label{d8b}
 \eeqn 

The matrixes $\hat M_i$ can be represented in the following form
using relations (\ref{d7a}), (\ref{d8b}),
 \beqn
\hat M_1 &=& 
\alpha_1\,\hat\nu_1(\vec k_1,\alpha_1)\,
\Bigl[\hat\mu_1(\vec k_1,\vec k_{23},\alpha_1,\alpha_2,\alpha_3)
- \hat\mu_1(\vec k_1,\vec k_{23} - \alpha_3\vec k_T,
\alpha_1,\alpha_2,\alpha_3)\Bigr]\nonumber\\  &+&
\alpha_2\alpha_3\,\Bigl[
\hat\lambda_1(\vec k_1,\vec k_{23},\alpha_1,\alpha_2,\alpha_3)
\nonumber\\ &-& \hat\lambda_1(\vec k_1-\vec k_T,\vec 
k_{23},\alpha_1,\alpha_2,\alpha_3)\Bigr]\,
\hat\rho_1(\vec k_{23},\alpha_2,\alpha_3)\ ;
\label{d9a}\\
\hat M_2 &=& \alpha_2\,\Bigl[
\hat\mu_2(\vec k_2,\vec
k_{13},\alpha_2,\alpha_1,\alpha_3) - 
\hat\mu_2(\vec k_2,\vec k_{13} - 
\alpha_3\vec k_T,\alpha_1,\alpha_2,\alpha_3)
\Bigr]\,\hat\nu_2(\vec k_2,\alpha_2)\nonumber\\
&+& \alpha_1\alpha_3\,
\hat\rho_2(\vec k_{13},\alpha_1,\alpha_3)\,
\Bigl[ \hat\lambda_2(\vec k_2,\vec 
k_{13},\alpha_2,\alpha_1,\alpha_3)\nonumber\\ 
&-& \hat\lambda_2(\vec k_2-\vec k_T,\vec 
k_{13},\alpha_1,\alpha_2,\alpha_3)
\Bigr]
\label{d9b}\\
\hat M_3 &=& 
\alpha_1\,\hat\nu_1(\vec k_1,\alpha_1)\,
\Bigl[\hat\mu_1(\vec k_1,\vec k_{23} - 
\alpha_3\vec k_T,
\alpha_1,\alpha_2,\alpha_3)\nonumber\\
&-& \hat\mu_1(\vec k_1,\vec k_{23} + \alpha_2\vec k_T,
\alpha_1,\alpha_2,\alpha_3)\Bigr]\ ;  
\label{d9c}\\
\hat M_4 &=& \alpha_2\,\Bigl[
\hat\mu_2(\vec k_2,\vec k_{13}-\alpha_3\vec k_T,
\alpha_2,\alpha_1,\alpha_3)\nonumber\\ 
&-& \hat\mu_2(\vec k_2,\vec k_{13} + 
\alpha_1\vec k_T,\alpha_1,\alpha_2,\alpha_3)
\Bigr]\,\hat\nu_2(\vec k_2,\alpha_2)\ ;
\label{d9d}\\
\hat M_5 &=& \alpha_2\alpha_3\,\Bigl[
\hat\lambda_1(\vec k_1-\vec k_T,\vec 
k_{23},\alpha_1,\alpha_2,\alpha_3)
\nonumber\\ &-& 
\hat\lambda_1(\vec k_1-\alpha_1\vec k_T,\vec 
k_{23},\alpha_1,\alpha_2,\alpha_3)\Bigr]\,
\hat\rho_1(\vec k_{23},\alpha_2,\alpha_3)\ ; 
\label{d9e}\\
\hat M_6 &=& 
\alpha_1\alpha_3\,
\hat\rho_2(\vec k_{13},\alpha_1,\alpha_3)\,
\Bigl[\hat\lambda_2(\vec k_2-\vec k_T,\vec 
k_{13},\alpha_1,\alpha_2,\alpha_3)\nonumber\\ 
&-& \hat\lambda_2(\vec k_2-\alpha_2\vec k_T,\vec 
k_{13},\alpha_1,\alpha_2,\alpha_3)\Bigr]\ ;
\label{d9f}\\
\hat M_7 &=& 
\alpha_3\,\hat\phi(\vec k_3,\alpha_1,\alpha_2,\alpha_3)\,
\Bigl[\hat\omega(\vec k_3,\vec k_{12}+\alpha_1\vec k_T,
\alpha_1,\alpha_2,\alpha_3)\nonumber\\
&-& \hat\omega(\vec k_3,\vec k_{12} - \alpha_2\vec k_T,
\alpha_1,\alpha_2,\alpha_3)\Bigr]\ ;
\label{d9g}\\ 
\hat M_8 &=& {1\over2}\, 
\alpha_3\,\hat\phi(\vec k_3,\alpha_1,\alpha_2,\alpha_3)\,
\Bigl[\hat\omega(\vec k_3,\vec k_{12}+\alpha_1\vec k_T,
\alpha_1,\alpha_2,\alpha_3)\nonumber\\ &+&
\hat\omega(\vec k_3,\vec k_{12}-\alpha_2\vec k_T,
\alpha_1,\alpha_2,\alpha_3)
- 2\,\hat\omega(\vec k_3,\vec k_{12},
\alpha_1,\alpha_2,\alpha_3)\Bigr]\ ;
\label{d9h}\\ 
\hat M_9 &=& 
\Bigl[\hat\beta(\vec k_3,\vec k_{12},\alpha_1,\alpha_2,\alpha_3) -
\hat\beta(\vec k_3-\vec k_T,\vec 
k_{12},\alpha_1,\alpha_2,\alpha_3)\Bigr]\nonumber\\
&\times& 
\hat\gamma(\vec k_{12},\alpha_1,\alpha_2,\alpha_3)\ . 
\label{d9i}
 \eeqn 
 The following notations are used here,
 \beqn
\hat\nu_1(\vec k_1,\alpha_1) &=&
\frac{\hat U_1(\vec k_1,\alpha_1)}
{\Delta_0(\vec k_1,\alpha_1)}\ ;
\nonumber\\
\hat\nu_2(\vec k_2,\alpha_2) &=&
\frac{\hat U_2(\vec k_2,\alpha_2)}
{\Delta_0(\vec k_2,\alpha_2)}\ ;
\label{d10}
 \eeqn
 \beqn
\hat\mu_1(\vec k_1,\vec k_{23},\alpha_1,\alpha_2,\alpha_3)
&=& \frac{\hat V_1(\vec k_{23},\alpha_2,\alpha_3)}
{\Delta_1(\vec k_1,\vec k_{23},\alpha_1,\alpha_2,\alpha_3)}
\ ;\nonumber\\ 
\hat\mu_2(\vec k_2,\vec k_{13},\alpha_1,\alpha_2,\alpha_3)
&=& \frac{\hat V_2(\vec k_{13},\alpha_1,\alpha_3)}
{\Delta_1(\vec k_2,\vec k_{13},\alpha_1,\alpha_2,\alpha_3)}
\ ;
\label{d11}
 \eeqn
 \beqn
\hat\lambda_1(\vec k_1,\vec k_{23},\alpha_1,\alpha_2,\alpha_3)
&=& \frac{\hat U_1(\vec k_1,\alpha_1)}
{\Delta_1(\vec k_1,\vec k_{23},\alpha_1,\alpha_2,\alpha_3)}
\ ;\nonumber\\ 
\hat\lambda_2(\vec k_2,\vec k_{13},\alpha_1,\alpha_2,\alpha_3)
&=& \frac{\hat U_2(\vec k_2,\alpha_2)}
{\Delta_1(\vec k_2,\vec k_{13},\alpha_1,\alpha_2,\alpha_3)}
\ ;
\label{d12}
 \eeqn
 \beqn
\hat\rho_1(\vec k_{23},\alpha_2,\alpha_3) &=&
\frac{\hat V_1(\vec k_{23},\alpha_2,\alpha_3)}
{\Delta_2(\vec k_{23},\alpha_2,\alpha_3)}\ ;
\nonumber\\
\hat\rho_2(\vec k_{13},\alpha_1,\alpha_3) &=&
\frac{\hat V_2(\vec k_{13},\alpha_1,\alpha_3)}
{\Delta_2(\vec k_{13},\alpha_1,\alpha_3)}\ ;
\label{d13}
 \eeqn
 \beq
\hat\beta(\vec k_3,\vec k_{12},\alpha_1,\alpha_2,\alpha_3)
= \frac{\hat V_0(k_3)}{\Delta_4(\vec k_3,\vec 
k_{12},\alpha_1,\alpha_2,\alpha_3)}\ ;
\label{d13a} 
 \eeq
 \beq
\hat\gamma(\vec k_{12},\alpha_1,\alpha_2,\alpha_3)
= \frac{\hat U_0(k_{12},\alpha_1,\alpha_2)}
{\Delta_5(\vec k_{12},\alpha_1,\alpha_2,\alpha_3)}\ ;
\label{d13b} 
 \eeq
 \beq
\hat\omega(\vec k_3,\vec k_{12},\alpha_1,\alpha_2,\alpha_3)
= \frac{\hat U_0(k_{12},\alpha_1,\alpha_2)}
{\Delta_4(\vec k_3,\vec k_{12},\alpha_1,\alpha_2,\alpha_3)}\ ;
\label{d13c} 
 \eeq
 \beq
\hat\phi(\vec k_3,\vec k_{12},\alpha_1,\alpha_2)
= \frac{\hat V_0(k_3)}{\Delta_3(\vec k_3,\alpha_1,\alpha_3)}\ ;
\label{d13d} 
 \eeq

Apparently, the amplitude operators in Eqs.~(\ref{d9a})-(\ref{d9i}) 
vanish, in forward direction, $\vec k_T \to 0$, what guarantees infra-red
stability of the cross section of gluon radiation. 

The above relations are valid at any values of $\alpha_1,\ 
\alpha_2,\ \alpha_3$, however they substantially simplify 
if the momentum fraction carried by the radiated gluon is 
small, $\alpha_3 \ll 1$ what corresponds to th dominant configuration for 
a fluctuation $G\to \bar ccG$. In the limit $\alpha_3 \to 0$ the
amplitudes
$M_1,\ M_2,\ M_5,\ M_6,\ M_7$ and $M_8$ vanish. Also the operators
$\hat V_{1,2}$ do not depend any more on the spin matrix 
$\vec\sigma$ and become proportional $\hat V_0$ 
 \beqn
\hat V_1(\vec k_{23},\alpha_2,\alpha_3)
\Bigr|_{\alpha_3\to 0} &=&
- \alpha_2^2\,\hat V_0(\vec k)\ ;
\nonumber\\
\hat V_2(\vec k_{13},\alpha_1,\alpha_3)
\Bigr|_{\alpha_3\to 0} &=&
- \alpha_1^2\,\hat V_0(\vec k)\ ;
\nonumber\\
\hat V_1(\vec k_{23}+\vec \alpha_2\vec k_T,
\alpha_2,\alpha_3)\Bigr|_{\alpha_3\to 0} &=&
- \alpha_2^2\,\hat V_0(\vec k - \vec k_T)\ ;
\nonumber\\
\hat V_2(\vec k_{13}+\alpha_1\vec k_T,
\alpha_1,\alpha_3)\Bigr|_{\alpha_3\to 0} &=&
- \alpha_1^2\,\hat V_0(\vec k-\vec k_T)\ .
\label{a100}
 \eeqn
 Also the matrixes $\hat U_{1,2}\propto \hat U_0$,
 \beqn
\hat U_1(\vec k_1,\alpha_1)
\Bigr|_{\alpha_3\to 0} &=&  
\hat U_0(\vec k_{12}+\alpha_1\vec k_T,
\alpha_1,\alpha_2)\ ;\nonumber\\
\hat U_2(\vec k_2,\alpha_2)
\Bigr|_{\alpha_3\to 0} &=&  
\hat U_0(\vec k_{12}-\alpha_2\vec k_T,
\alpha_1,\alpha_2)\ .
\label{a110} 
 \eeqn
 Besides, $\alpha_1+\alpha_2=\bar\alpha_3 \approx 1$ and the 
denominators $\Delta_i$ take the simple form,
 \beqn 
\Delta_0(\vec k_1,\alpha_1)
\Bigr|_{\alpha_3\to 0} &=&
m_c^2+k_1^2-\alpha_1\alpha_2\,\lambda^2\ ;
\nonumber\\
\Delta_0(\vec k_2,\alpha_1)
\Bigr|_{\alpha_3\to 0} &=&
m_c^2+k_2^2-\alpha_1\alpha_2\,\lambda^2\ ;
\nonumber\\
\Delta_5(\vec k_{12},\alpha_1,\alpha_2,\alpha_3)
\Bigr|_{\alpha_3\to 0} &=& 
m_c^2+k_{12}^2-\alpha_1\alpha_2\,\lambda^2\ ;
\nonumber\\
\Delta_1(\vec k_1,\vec k_{23},\alpha_1,\alpha_2,\alpha_3)
\Bigr|_{\alpha_3\to 0} &=&
\alpha_2\alpha_1^2\,
\Bigl[k_3^2+\tau^2(k_1^2)\Bigr]\ ;
\nonumber\\
\Delta_1(\vec k_2,\vec k_{13},\alpha_1,\alpha_2,\alpha_3)
\Bigr|_{\alpha_3\to 0} &=&
\alpha_1\alpha_2^2\,
\Bigl[k_3^2+\tau^2(k_2^2)\Bigr]\ ;
\nonumber\\
\Delta_4(\vec k_3,\vec k_{12},\alpha_1,\alpha_2,\alpha_3)
\Bigr|_{\alpha_3\to 0} &=&
\alpha_1\alpha_2\,
\Bigl[k_3^2+\tau^2(k_{12}^2)\Bigr]\ ,
\label{a120}  
 \eeqn  
 where
 \beq
\tau^2(\vec k) = \frac{\alpha_3}{\alpha_1\alpha_2}\,
(m_c^2+k^2) + \lambda^2\ .
\label{a130}
 \eeq
 
We keep the term linear in $\alpha_3$ in (\ref{a130}) since it is responsible
for suppression of gluon radiation with 
 \beq
\alpha_3 \gsim \frac{\lambda^2}{M^2_{\bar cc}}\ ,
\label{a140}
 \eeq
where
 \beq
M^2_{\bar cc} = \frac{m_c^2+k_{2}^2}
{\alpha\bar\alpha}
\label{a144}
 \eeq
is the invariant mass squared of the $\bar cc$ pair, and
 \beq
\vec k_{12} = \bar\alpha\vec k_1 - 
\alpha\vec k_2\ .
\label{a146}
 \eeq

The approximation $\alpha_3\ll 1$ is justified even for gluons satisfying the
condition Eq.~(\ref{a140}), since
 \beq
\frac{\lambda^2}{M^2_{\bar cc}} \sim
\frac{\Lambda_{QCD}^2}{4\,m_c^2} \sim
0.005\ .
\label{a150}
 \eeq

Further, we will make use of the relations,
 \beqn
\vec k_1 &=& \vec k_{12} +
\alpha(\vec k_T - \vec k_3) +
O(\alpha_3)\nonumber\\
\vec k_2 &=& \vec k_{12} +
\bar\alpha(\vec k_T - \vec k_3) +
O(\alpha_3)\ ,
\label{a160}
 \eeqn
where $\alpha=\alpha_1$, and also take into account that both
$k_T$ and $k_3$ are of the order of $\lambda$, i.e. they are much smaller than
the typical values of $k_{12}\sim m_c$. Therefore, we will rely on the
relations,
 \beq
\tau^2(k_1)\approx \tau^2(k_2) \approx
 \tau^2(k_{12}) \approx \tau^2 = 
\lambda^2 +\alpha_3\,M_{\bar cc}^2\ .
\label{a170}
 \eeq
Within this approximation the amplitude of reaction
$GN\to\bar cc\,G\,X$ at $\alpha_3\ll1$ is given by
 \beqn
&& A_{ij,ab}^{\bar\mu\mu} =
i\alpha_s^{3/2}\sum\limits_{e,d=1}^{N_c^2-1}
\frac{f_{bde}\,F^d_{GN\to X}(\vec k_T,\{x\})}
{k_T^2+\lambda^2}\
\xi^\mu_c\Bigr.^\dagger\left\{
\frac{\hat U_0(\vec k_1,\alpha,\bar\alpha)}
{k_1^2+m_c^2}\,(\tau_e\,\tau_a)_{ij} 
\right.\nonumber\\ &-& \left.
\frac{\hat U_0(-\vec k_2,\alpha,\bar\alpha)}
{k_2^2+m_c^2}\,(\tau_a\,\tau_e)_{ij} +
i\,\frac{\hat U_0(\vec k_{12},\alpha,\bar\alpha)}
{k_{12}^2+m_c^2}\,
\sum\limits_{g=1}^{N_c^2-1} f_{aeg}\,(\tau_g)_{ij}
\right\}\tilde\xi_{\bar c}^{\bar\mu}
\nonumber\\ &\times&\left[
\frac{2\vec e_f\cdot(\vec k_3-\vec k_T)}
{(\vec k_3-\vec k_T)^2 + \tau^2} -
\frac{2\,\vec e_f\cdot\vec k_3}
{\vec k_3^2 + \tau^2}\right]\ .
\label{a180}
\eeqn

To switch to the impact parameter representation we should perform Fourier
integration over momenta $\vec k_{12},\ \vec k_T$ and $\vec k_3$.  This
procedure turns out to be quite complicated because of the dependence of
$\tau$ in (\ref{a170}) on the invariant mass Eq.~(\ref{a144})  which
involves $\vec k_{12}$. For the sake of simplicity we replace
in Eqs.~(\ref{a170}) -- (\ref{a180}) the invariant mass of the $\bar cc$
pair by its mean value, $M_{\bar cc}^2 \Rightarrow \overline{M_{\bar
cc}^2} = N\,4m_c^2$, where $N\sim 1-2$. Then Eq.~(\ref{a180}) can be
represented in the form,
 \beqn
&& A_{ij,ab}^{\bar\mu\mu} =
3i\sum\limits_{e,d=1}^{N_c^2-1} f_{bde}
\int d^2r\,d^2\rho\,d^2s\,
\exp\Bigl(i\,\vec k_{12}\cdot\vec r +
i\,\vec k_3\cdot\vec\rho +
i\,\vec k_T\cdot\vec s\Bigr)
\nonumber\\ &\times&
\Psi^{\bar\mu\mu}(\vec r,\alpha)\,
\Biggl\{
\Phi_{cG}(\vec\rho+\alpha\vec r)\ (\tau_a\tau_e)_{ij}\,
\Bigl[\gamma^{(d)}(\vec\rho+\vec s) - 
\gamma^{(d)}(\vec s-\alpha\vec r)\Bigr]
\nonumber\\ &-& \left.
\Phi_{cG}\Bigl(\vec\rho-\bar\alpha\vec r\Bigr)
\ (\tau_e\tau_a)_{ij}\,
\Bigl[\gamma^{(d)}(\vec\rho+\vec s) -
\gamma^{(d)}\Bigl(\vec s+\bar\alpha\vec r\Bigr)\Bigr]
\right.\nonumber\\ &+& 
\Phi_{cG}(\vec\rho)\,\sum\limits_{g=1}f_{eag}\ 
(\tau_g)_{ij}\,
\Bigl[\gamma^{(d)}(\vec\rho+\vec s) -
\gamma^{(d)}(\vec s)\Bigr]\ ,
\Biggr\}
\label{a190}
 \eeqn
 where $\vec r$, $\vec\rho$ and $\vec s$ are the inter-quark separation,
gluon -- $\bar cc$ separation, and the position of the center of gravity
of the whole $\bar ccG$ system, respectively.  The $\bar cc$ LC wave
function $\Psi^{\bar\mu\mu}(\vec r,\alpha)$ and the profile function
$\gamma^{(d)}(\vec s)$ are defined in (\ref{370}) and (\ref{380}),
respectively, and
 \beq
\Phi_{cG}(\vec\rho) = 
\frac{2i\sqrt{\alpha_s}}
{\pi\sqrt{3}}\ 
\vec e_f\cdot\vec\nabla_\rho\ 
K_0(\tau\rho)\ .
\label{a200}
 \eeq

Further, note that the mean $\bar cc$ separation $\sim 1/m_c$ is much
smaller than the typical scale $\sim 1/\lambda\sim 1/\Lambda_{QCD}\sim
1\,\fm$ of variation of the profile function $\gamma^{(d)}(s)$. Therefore,
we can neglect the $\vec r$-dependence of $\gamma^{(d)}$ in (\ref{a190}),
i.e. replace
 \beqn
\gamma^{(d)}(\vec s-\alpha r) 
&\Rightarrow&
\gamma^{(d)}(\vec s)\ ,
\nonumber\\
\gamma^{(d)}\Bigl(\vec s+\bar\alpha r\Bigr)    
&\Rightarrow&
\gamma^{(d)}(\vec s)\ .
\label{a210}
 \eeqn

After replacing the bilinear to linear combinations of $\tau$-matrixes in
(\ref{a190}) by means of relation
 \beq
(\tau_a\tau_e)_{ij} = {1\over6}\,\delta_{ae}\,\delta_{ij} +
{1\over2}\sum\limits_{g=1}^{N_c^2-1}
(d_{aeg}+if_{aeg})\,(\tau_g)_{ij}\ ,
\label{a220}
 \eeq
we arrive at the final result of this section,
 \beqn
&& A_{ij,ab}^{\bar\mu\mu} =
3i\sum\limits_{e,d=1}^{N_c^2-1} f_{bde}
\int d^2r\,d^2\rho\,d^2s\,
\exp\Bigl(i\,\vec k_{12}\cdot\vec r +
i\,\vec k_3\cdot\vec\rho +
i\,\vec k_T\cdot\vec s\Bigr)\,
\Bigl[\gamma^{(d)}(\vec\rho+\vec s)-
\gamma^{(d)}(\vec s)\Bigr]
\nonumber\\ &\times& \Psi^{\bar\mu\mu}(\vec r,\alpha)\,
\Biggl\{\left[\,{1\over6}\,\delta_{ae}\delta_{ij}+
{1\over2}\sum\limits_g d_{aeg}(\tau_g)_{ij}\,\right]\,
\Phi^-_{cG}(\vec r,\vec\rho)+
{i\over2}\sum\limits_g f_{aeg}(\tau_g)_{ij}\,
\Phi^+_{cG}(\vec r,\vec\rho)\Biggr\}\ .
\label{a230} 
 \eeqn
 Here
 \beqn
\Phi^-_{cG}(\vec r,\vec\rho) &=& 
\Phi_{cG}(\vec\rho+\alpha\,\vec r) -
\Phi^-_{cG}\Bigl(\vec\rho - \bar\alpha\vec r\Bigr)\ ;
\nonumber\\
\Phi^+_{cG}(\vec r,\vec\rho) &=& 
\Phi_{cG}(\vec\rho+\alpha\,\vec r) +
\Phi^-_{cG}\Bigl(\vec\rho - \bar\alpha\vec r\Bigr)
- 2\,\Phi_{cG}(\vec\rho)\ .
\label{a240}
 \eeqn

The three terms in curly brackets in Eq.~(\ref{a230}) describe production
of the $\bar cc$ pair in different color and spin states. The first term
containing $\delta_{ae}\delta_{ij}$ is responsible for production of a
colorless $\bar cc$ with positive $C$-parity. Since $C$-transformation
cannot be applied to colored states, instead we use parity relative to
interchange of the $c$ and $\bar c$ quarks., which is negative for this
state. Therefore we classify this colorless state as $1^-$ ($1$ means
color singlet).
The second and third terms in (\ref{a230}) containing $\sum\limits_g
d_{aeg}(\tau_g)_{ij}$ and $\sum\limits_g f_{aeg}(\tau_g)_{ij}$ correspond
to production of color-octet $\bar cc$, $8^-$ and $8^+$ respectively.

Note that the $\bar cc$ pair can be also produced in a colorless state
$1^+$, i.e. with negative $C$-parity. This is not forbidden in the process
$GN\to \bar ccGN$ by selection rules, like it happens in reaction
$GN\to \bar ccN$, but is suppressed dynamically. Indeed, it turns out
that the production amplitude is proportional to $\alpha_G$, therefore it
can be neglected in the limit of $\alpha_G \ll 1$ we are interested in. 

As it was first pointed out in \cite{kst2} the nonperturbative interaction
between the gluon and the $\bar cc$ pair in the $|\bar ccG\ra$ Fock state
may significantly modify the LC wave function of this state. Therefore,
one should replace the perturbative wave functions Eq.~(\ref{a200})  by
the nonperturbative ones \cite{kst2}. Apparently, the result will depend
on the strength of the LC potential which should be different for the
three different states of the $\bar cc$ pair discussed above. 

According to the approximation accepted above that the distance between the
gluon and the center of gravity of the $\bar cc$ pair is much larger than
the the pair itself, the combinations Eq.~(\ref{a240}) of the wave
functions may be approximated as,
 \beqn
\Phi^-_{cG}(\vec r,\vec\rho) &\approx&
\vec r\cdot\vec\nabla_\rho\,
\Phi_{cG}(\vec\rho)\ ;
\nonumber\\
\Phi^+_{cG}(\vec r,\vec\rho) &\approx&
(2\alpha-1)\,\vec r\cdot\vec\nabla_\rho\,
\Phi_{cG}(\vec\rho)\ .
\label{a250}
 \eeqn



\begin{thebibliography}{99}

\bibitem{mike} E789 Collaboration, M.J.~Leitch et al., Phys. Rev. Lett.
{\bf 72} (1994) 2542.

\bibitem{e769} E769 Collaboration, G.A.~Alves et al., Phys. Rev. Lett.
{\bf 70} (1993) 722.

\bibitem{wa82} WA82 Collaboration, M. Adamovich et al., Phys. Lett. B{\bf
284} (1992) 453. 

\bibitem{nmc} NMC Coll., M.~Arneodo et al., Nucl. Phys. B{\bf 481}
(1996) 23.

\bibitem{e772} The E772 Collaboration, D.M.~Alde et al, Phys. Rev.  Lett.
{\bf 64} (1990) 2479.

\bibitem{kth} B.Z. Kopeliovich, A.V. Tarasov, J.~H\"ufner,
Nucl. Phys. A{\bf 696} (2001) 669

\bibitem{eks} K. J. Eskola, V. J. Kolhinen, and P. V. Ruuskanen, Nucl.
Phys. {\bf B535} (1998) 351; K. J. Eskola, V. J. Kolhinen, and
C.A.~Salgado, Eur. Phys. J. {\bf C9} (1999) 61.

\bibitem{kumano}  M. Hirai, S. Kumano, M. Miyama, Phys. Rev. D{\bf 64} (2001)
034003

\bibitem{eks-new} K.J. Eskola, H. Honkanen, V.J. Kolhinen, C.A. Salgado,
hep-ph/0201256

\bibitem{hir} B.Z.~Kopeliovich, {\sl Dynamics and Phenomenology of Charmonium
Production off Nuclei}, in proc. of the Workshop Hirschegg'97: QCD Phase
Transitions', Hirschegg, Austria, January, 1997, ed. by H.~Feldmeier,
J.~Knoll, W.~N\"orenberg and J.~Wambach, Darmstadt, 1997, p. 281;
hep-ph/9702365

\bibitem{krtj}  B.Z. Kopeliovich, J. Raufeisen, A.V. Tarasov, M.B.~Johnson,
hep-ph/0110221.

\bibitem{gay} B.Z.~Kopeliovich, J.~Raufeisen and A.V.~Tarasov,
Phys. Lett. B {\bf 503} (2001) 91;
M.A. Betemps, M.B. Gay Ducati, M.V.T. Machado, hep-ph/0111473. 

\bibitem{zkl} A.B.~Zamolodchikov, B.Z.~Kopeliovich and L.I.~Lapidus,
Sov. Phys. JETP Lett. {\bf 33} (1981) 612.

\bibitem{joerg} J. Raufeisen, J.-C. Peng, G.C. Nayak, hep-ph/0204095. 

\bibitem{bhq}  S.J.~Brodsky, A.~Hebecker and  E.~Quack,
 Phys. Rev. {\bf D55} (1997) 2584.

\bibitem{kst1} B.Z.~Kopeliovich, A.~Sch\"afer and A.V.~Tarasov,
Phys. Rev. {\bf C59} (1999) 1609 (extended version in hep-ph/9808378)
 
\bibitem{kst2} B.Z.~Kopeliovich, A.~Sch\"afer and A.V.~Tarasov, Phys. Rev. {\bf
D62} (2000) 054022

\bibitem{jkt} M.B. Johnson, B.Z.~Kopeliovich, A.V.~Tarasov, Phys. Rev. C{\bf
63} (2001) 035203. 

\bibitem{mv} L.~McLerran and R.~Venugopalan, Phys. Rev. D {\bf 49}, 2233
(1994); {\bf 49}, 3352 (1994); {\bf 49}, 2225 (1994).

\bibitem{al} A.H. Mueller, {\it Parton saturation: an overview},
hep-ph/0111244

\bibitem{npz} N.N. Nikolaev, G.~Piller, B.G. Zakharov, JETP {\bf 81} (1995)
851 [Zh. Eksp. Teor. Fiz. {\bf 108} (1995) 1554. 

\bibitem{lp} L.D.Landau, I.Ya.Pomeranchuk, {\it ZhETF} {\bf 24} (1953) 505,
\\ L.D.Landau, I.Ya.Pomeranchuk, {\it Doklady AN SSSR} {\bf 92} (1953) 535,
735\\ E.L.Feinberg, I.Ya.Pomeranchuk, {\it Doklady AN SSSR} {\bf 93} (1953)
439, \\ I.Ya.Pomeranchuk, {\it Doklady AN SSSR} {\bf 96} (1954) 265, \\
I.Ya.Pomeranchuk, {\it Doklady AN SSSR} {\bf 96} (1954) 481, \\ E.L.Feinberg,
I.Ya.Pomeranchuk, {\it Nuovo Cim. Suppl.} {\bf 4} (1956) 652. 

\bibitem{mueller} A.H.~Mueller, Nucl. Phys. {\bf B335} (1990) 115;
{\it ibid} {\bf B558} (1999) 285.

\bibitem{krt1} B.Z.~Kopeliovich, J.~Raufeisen and A.V.~Tarasov,
Phys. Lett. {\bf B440} (1998) 151; J.~Raufeisen, A.V.~Tarasov and
O.O.~Voskresenskaya, Eur. Phys. J. A {\bf 5} (1999) 173.
 
\bibitem{krt2}  B.Z.~Kopeliovich, J.~Raufeisen and A.V.~Tarasov,
Phys. Rev. {\bf C62} (2000) 035204.

\bibitem{knst} B.Z. Kopeliovich, J. Nemchik, A. Sch\"afer, A.V.~Tarasov,
Phys. Rev. C{\bf 65} (2002) 035201.

\bibitem{feynman} R.P.~Feynman and A.R.~Gibbs, {\sl Quantum Mechanics and
Path Integrals}, McGRAW--HILL Book Company, New York 1965

\bibitem{k3p} B.Z.~Kopeliovich, I.K.~Potashnikova, B.~Povh and E.~Predazzi,
Phys. Rev. Lett. {\bf 85} (2000) 507; Phys. Rev. D{\bf 63} (2001) 054001. 

\bibitem{pisa} M.~D'Elia, A.~Di~Giacomo and E.~Meggiolaro,
Phys. Lett. {\bf B408} (1997) 315.

\bibitem{shuryak} T.~Sch\"afer, E.V.~Shuryak, Rev. Mod. Phys.
{\bf 70} (1998) 323.

\bibitem{andreas} V.M.~Braun, P.~G\'ornicki, l.~Mankiewicz
and A.~Sch\"afer, Phys. Lett. {\bf B302} (1993) 291.

\bibitem{km} Yu.V.~Kovchegov and A.H.~Mueller, Nucl.  Phys. B{\bf 529},
451 (1998).

\bibitem{dima} R. Gavai et al., Int. J. Mod. Phys. A{\bf 10} (1995) 3043.

\bibitem{chi} E705 Collaboration, L.~Antoniazzi et al., Phys. Rev. Lett. 
{\bf 70} (1993) 383

\bibitem{andreas1} Ph. Hagler, R. Kirschner, A.~Sch\"afer, L.~Szymanowski,
O.V.~Teryaev, Phys. Rev. Lett. {\bf 86} (2001) 1446.

\bibitem{chi1-chi2} W11 Collaboration, Y.~Lemoigne et al., Phys. Lett. B{\bf
113} (1982) 509.

\bibitem{baier} R.~Baier and R.~R\"uckl, Z. Phys. {\bf C19} (1983) 251

\bibitem{brodsky} M. Vanttinen, P. Hoyer, S.J.~Brodsky, Wai-Keung~Tang,
Phys. Rev. {\bf D51} (1995) 332

\bibitem{michele} M.~Arneodo, Phys. Rep. {\bf 240} (1994) 301.

\bibitem{eloss} M.B.~Johnson et al., Phys.Rev.Lett. {\bf 86}, 4483
(2001); Phys. Rev. C {\bf 65}, 025203 (2002).

\bibitem{phenix-charm} PHENIX Coll., K.~Adcox et al., nucl-ex/0202002.

\bibitem{hikt} J. H\"ufner, Yu.P.~Ivanov, B.Z.~Kopeliovich, A.V.~Tarasov,
Phys. Rev. D{\bf 62} (2000) 094022. 

\bibitem{gbw} K.~Golec-Biernat and M.~W\"usthoff, Phys. Rev. D{\bf 59} (1999)
014017. 

\bibitem{ekv} K.J.~Eskola, V.J.~Kolhinen, R.Vogt, Nucl. Phys. A{\bf 696}
(2001) 729. 

\end{thebibliography}
\end{document}